\documentclass[12pt,a4paper,superscriptaddress,preprint]{revtex4}
\usepackage[dvips]{graphicx}
\usepackage{amssymb,amsmath}
\usepackage{color}
\begin{document}

\newcommand{\EQ}{Eq.~}
\newcommand{\EQS}{Eqs.~}
\newcommand{\FIG}{Fig.~}
\newcommand{\FIGS}{Figs.~}
\newcommand{\TAB}{Tab.~}
\newcommand{\TABS}{Tabs.~}
\newcommand{\SEC}{Sec.~}
\newcommand{\SECS}{Secs.~}

\title{Analysis of relative influence of nodes in directed networks}
\author{Naoki Masuda}
\affiliation{Graduate School of Information Science and Technology,
The University of Tokyo,
7-3-1 Hongo, Bunkyo, Tokyo 113-8656, Japan}
\affiliation{PRESTO, Japan Science and Technology Agency,
4-1-8 Honcho, Kawaguchi, Saitama 332-0012, Japan}
\author{Yoji Kawamura}
\affiliation{Institute for Research on Earth Evolution,
Japan Agency for Marine-Earth Science and Technology,
3173-25 Showa-machi, Kanazawa-ku, Yokohama, Kanagawa 236-0001, Japan}
\author{Hiroshi Kori}
\affiliation{Division of Advanced Sciences, Ochadai Academic Production,
Ochanomizu University,
2-1-1, Ohtsuka, Bunkyo-ku, Tokyo 112-8610, Japan}
\affiliation{PRESTO, Japan Science and Technology Agency,
4-1-8 Honcho, Kawaguchi, Saitama 332-0012, Japan}

\begin{abstract}
Many complex networks are described by directed links; in such networks, a link represents, for example, the control of one node over the other node or unidirectional information flows. Some centrality measures are used to determine the relative importance of nodes specifically in directed networks. We analyze such a centrality measure called the influence. The influence represents the importance of nodes in various dynamics such as synchronization, evolutionary dynamics, random walk, and social dynamics. We analytically calculate the influence in various networks, including directed multipartite networks and a directed version of the Watts-Strogatz small-world network. The global properties of networks such as hierarchy and position of shortcuts, rather than local properties of the nodes, such as the degree, are shown to be the chief determinants of the influence of nodes in many cases. The developed method is also applicable to the calculation of the PageRank.  We also numerically show that in a coupled oscillator system, the threshold for entrainment by a pacemaker is low when the pacemaker is placed on influential nodes. For a type of random network, the analytically derived threshold is approximately equal to the inverse of the influence. We numerically show that this relationship also holds true in a random scale-free network and a neural network.
\end{abstract}

\maketitle

\newpage

\section{Introduction}\label{sec:introduction}

Networks abound in various fields; a network is a
collection of nodes and links, where a link connects a pair of nodes.
Most real-world networks are not entirely regular or random
and have prominent properties as modeled by, for example,
small-world, scale-free, hierarchical, and modular networks
\cite{reviews,Boccaletti06}. In such networks, some nodes are
considered to be more important than the
others. Depending on the definition of importance, various centrality measures, which quantify the relative
importance of different nodes, have been proposed. The
most frequently used centrality measures are perhaps the degree (i.e., the
number of links owned by a node) and the betweenness (i.e., the
normalized number of shortest paths connecting any pair of nodes
passing through the
node in question) \cite{Boccaletti06,Freeman79,Wasserman94}.
New centrality measures have
also been proposed in the field of complex networks
\cite{Noh04prl-Newman05soc-Estrada05-Latora07,Restrepo06prl}.

Although many centrality measures are available, very few of
these describe the importance of nodes in collective
behavior of nodes on networks
(see \cite{Restrepo06prl}).  In a
previous study, we proposed a centrality measure called 
the influence \cite{MKK09-1}.
The influence of a node denotes its importance in different
types of dynamics.  It represents the amplitude of the
response of a synchronized network
when an input is given to a certain node \cite{Kori08}, the
fixation probability for a newly introduced type 
(e.g., new information) at a node in
voter-type evolutionary dynamics \cite{Masuda09njp}, the stationary
density of a simple random walk in continuous time
\cite{Masuda09njp}, the so-called reproductive value of a node
\cite{Taylor90-Taylor96}, and the influence of a node in
the DeGroot's model of consensus formation \cite{Degroot74-Friedkin91}.
It makes sense to consider the 
influence only in directed networks; in undirected networks,
the influences of all the nodes take an identical value.
In principle, the influence as a centrality measure
is close to the PageRank, which was originally developed
for ranking websites \cite{Brin98-Berkhin05}.

To assess the influence (and also the PageRank) in real complex
networks, it is not sufficient to take into account
the local property of the
node, such as the degree. The global structure of networks such as the
small-world property, modular structure, and self-similarity
\cite{reviews,Boccaletti06} generally affects the influence values.

In the present study, we analytically determine the influence of nodes
in model networks such as weighted chain, directed multipartite
networks with a hierarchical structure, and a directed version of the
Watts-Strogatz small-world network \cite{Watts98}. For this purpose,
we exploit the symmetry in networks and the relationship between the
enumeration of directed spanning trees and the influence. We reveal
the discrepancy between the actual influence and that predicted by the
mean-field approximation (MA), which takes into account only the
degree.  In fact, the nodes that occupy globally important positions
in terms of influence are generally different from those that are
locally important.  The globally important nodes govern the
above-mentioned dynamics on networks.  Finally, to demonstrate the
application of the influence as a centrality measure, we analyze a
system of coupled oscillators and show that nodes with large influence
values entrain other nodes relatively easily, i.e., with a relatively small
coupling strength.

\section{Influence}

\begin{figure}
\begin{center}
\includegraphics[width=6cm]{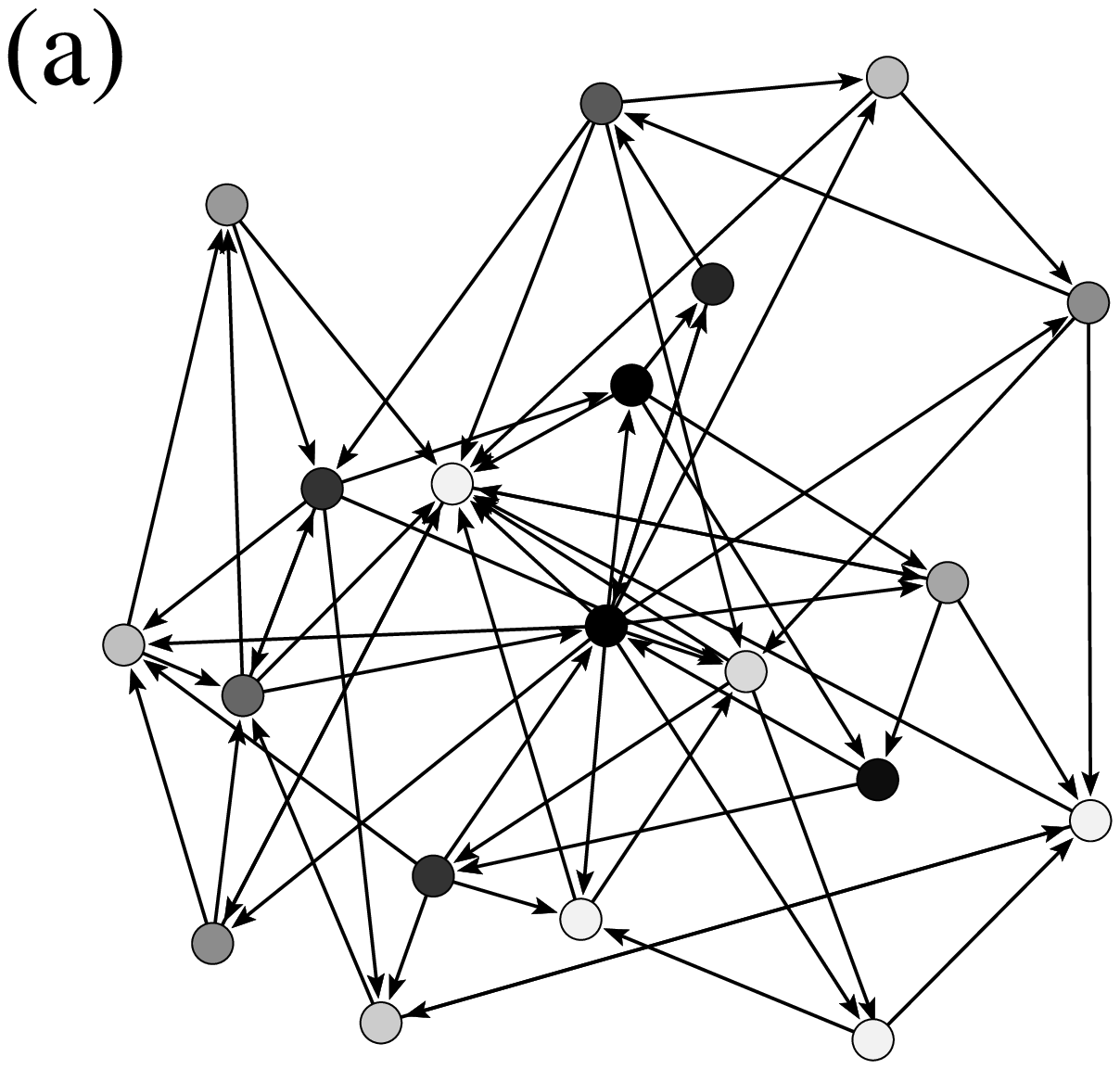}
\includegraphics[width=6cm]{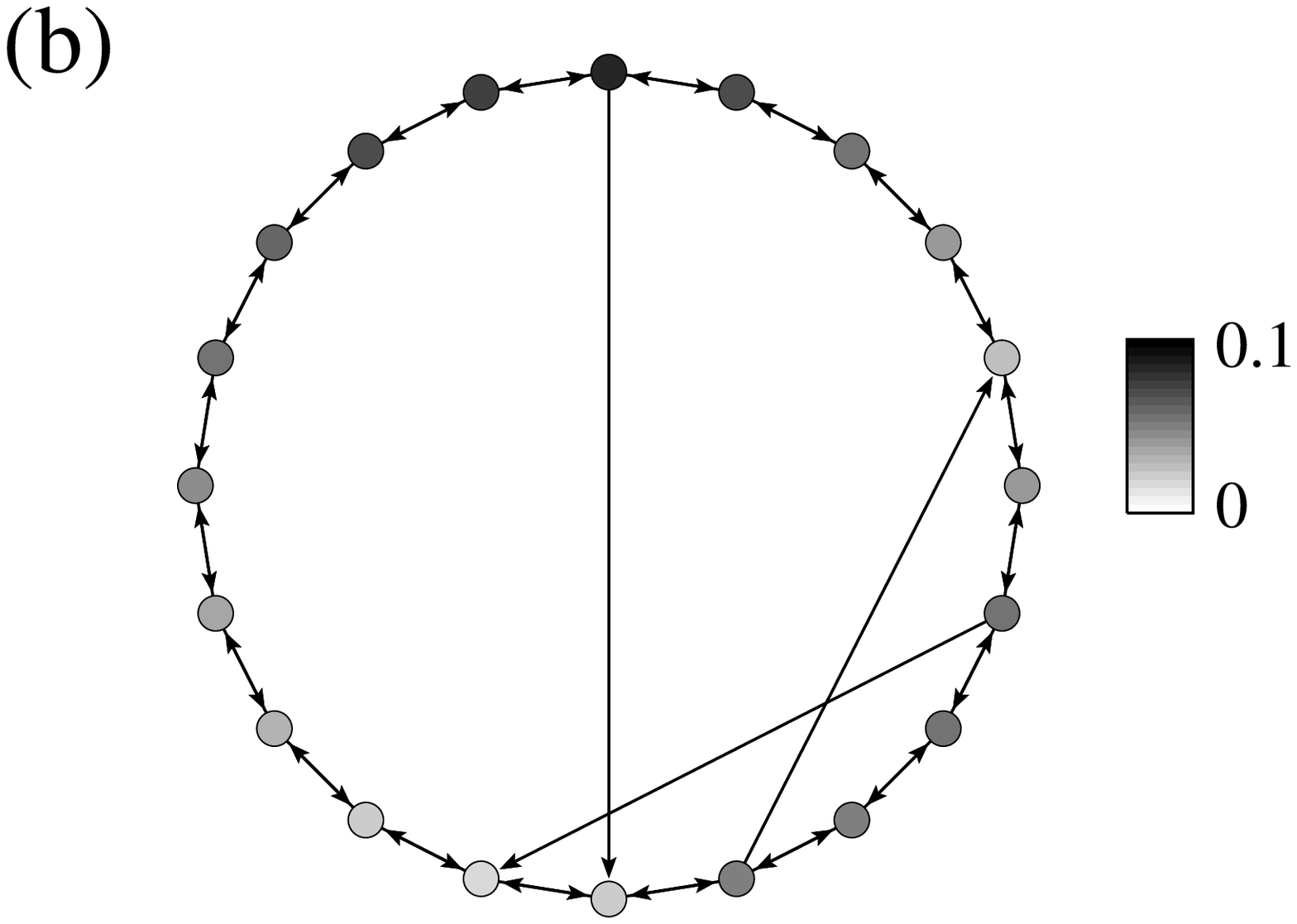}
\caption{Influence of nodes in networks having $N=20$.  A dark
node has a large value of $v_i$.  (a) Directed network
generated by the configuration model.  The indegree and outdegree
follow independent power-law distributions with the scaling exponent
2.5 and the minimum degree 2.  (b) Directed Watts-Strogatz network
with three shortcuts.  See \SECS\ref{sub:sw} and \ref{sec:kcr} for
details of the network models. The networks are visualized by Pajek
\cite{pajek}.}
\label{fig:example nets}
\end{center}
\end{figure}

Consider a directed and weighted
network having $N$ nodes.
The weight of the directed link from node $i$ to node $j$ is denoted by
$w_{ij}$. We set $w_{ij}=0$ when the link is absent.
The influence of node $i$ is denoted by $v_i$.
We define $v_i$ as the solution for
the following set of $N$ linear equations:
\begin{equation}
v_i = \frac{\sum_{j=1}^N w_{ij}v_j}{k_i^{\rm in}},\quad (1\le i\le N),
\label{eq:influence}
\end{equation}
where $k_i^{\rm in}\equiv \sum_{j=1}^N w_{ji}$ is the indegree of node
$i$, and the normalization is given by $\sum_{i=1}^N v_i=1$.
When node $i$ has many outgoing links, $v_i$ can be large because
there are many terms on the right-hand side of
\EQ\eqref{eq:influence}.  When node $i$ has many incoming links,
node $i$ is interpreted to be governed by many nodes.
Then, $v_i$
can be small because of the divisive factor $k_i^{\rm in}$ in
\EQ\eqref{eq:influence}. The rationale for
the definition given in \EQ\eqref{eq:influence} is that
$v_i$ represents the importance of nodes in
different types of dynamics on networks, as explained in
\SEC\ref{sec:introduction}. Values of $v_i$ for two example networks
are shown in \FIG\ref{fig:example nets}.

Note that $v_i=1/N$ for any network with $k_i^{\rm in}=k_i^{\rm out}$ 
($1\le i\le N$), where 
$k_i^{\rm out}\equiv \sum_{j=1}^N w_{ij}$ is the outdegree.
The undirected network is included in this class of networks.
Therefore, the influence has a nontrivial meaning only in
directed networks. This situation also holds true in the case of
 the PageRank; in undirected networks,
the PageRank is a linear function of
the degree of node \cite{Brin98-Berkhin05}.

The MA of $v_i$ is given by
\begin{equation}
v_i = \frac{\sum_{j=1}^N w_{ij}v_j}{k_i^{\rm in}}
\approx \frac{\sum_{j=1}^N w_{ij}\bar{v}}{k_i^{\rm in}}
\propto \frac{k_i^{\rm out}}{k_i^{\rm in}},
\label{eq:influence MA}
\end{equation}
where $\bar{v}\equiv \sum_{j=1}^N v_j/N=1/N$.
In \SECS\ref{sec:examples} and \ref{sec:kcr},
we argue that \EQ\eqref{eq:influence MA}
does not satisfactorily describe
$v_i$ in certain practically important types of networks,
including the Watts-Strogatz small-world network.
For these networks, we calculate the exact $v_i$ by using
different methods. 

The value of $v_i$ can be associated with the number of
directed spanning trees rooted at node $i$, as described below.
Equation~\eqref{eq:influence} implies that
$v_i$ is the left eigenvector
of the Laplacian matrix $L$, whose ($i,j$) element is equal to
$L_{ii}=\sum_{j=1,j\neq i}^N w_{ji}$ and $L_{ij}=-w_{ji}$ ($i\neq j$).
The corresponding eigenvalue is equal to 0; 
$\sum^N_{i=1}v_iL_{ij}=0$ ($1\le j\le N$). 
The $(i,j)$ cofactor of $L$ is given as
\begin{equation}
D\left( i, j \right)
\equiv (-1)^{i+j} \det L\left( i, j \right),
\label{eq:adjoint}
\end{equation}
where $L(i,j)$ is an $(N-1)\times (N-1)$ matrix obtained by deleting
the $i$th row and the $j$th column of $L$.
Because $\sum_{j=1}^N L_{ij} = 0$ ($1\le i\le N$),
$D\left( i, j \right)$ is independent of $j$.
Therefore, considering the fact that
$L$ has eigenvalue 0, we obtain
\begin{align}
  \sum_{i=1}^N D(i,i) L_{ij}
  &= \sum_{i=1}^N D(i,j)L_{ij}\nonumber\\
  &= \det L = 0,\quad (1\le j\le N).
\label{eq:det(L)=0}
\end{align}
Equation~\eqref{eq:det(L)=0} indicates that
$\left[D\left( 1, 1 \right),
\ldots, D\left( N, N \right)\right]$ is the left eigenvector
of $L$ with eigenvalue 0. Therefore, we obtain
\begin{equation}
v_i\propto D\left( i,i\right) 
=\det L\left( i,i\right).
\end{equation}

According to the matrix tree theorem \cite{ref:biggs-ref:agaev},
$\det L\left( i,i\right)$
is equal to the sum of the weights of 
all directed spanning trees of $G$ rooted at node $i$. The
weight of a spanning tree is defined as the product of 
the weights of the $N-1$ links used in the spanning tree.
Therefore, we can calculate $v_i$ by enumerating the spanning trees.

\section{Calculation of influence in some model networks}
\label{sec:examples}

In this section, we analytically calculate the influence for several
networks. Through these calculations, we show that
the influence extracts the globally important nodes, 
which is beyond the scope of the MA: $k_i^{\rm out}/k_i^{\rm in}$.
Such nodes are located upstream in the
hierarchy that is defined by the directionality of the links or
around the source of a valuable directed shortcut.

\subsection{Weighted chain}\label{sub:chain}

\begin{figure}
\begin{center}
\includegraphics[width=6cm]{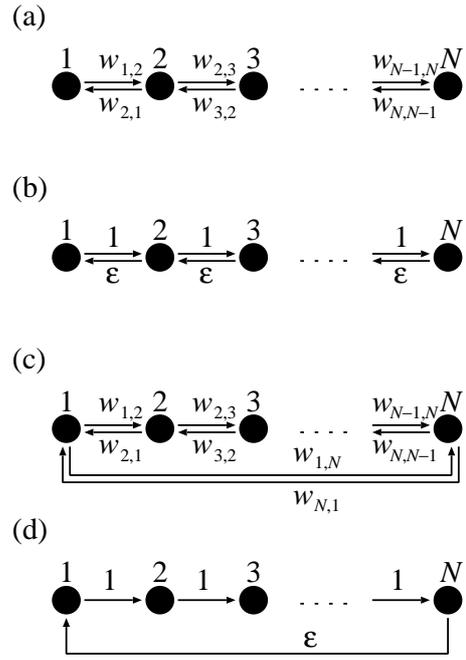}
\caption{Schematic of (a) weighted chain, (b) special case of weighted
chain, (c) weighted cycle, and (d) special case of weighted cycle.}
\label{fig:schem dim1}
\end{center}
\end{figure}

Consider a weighted 
chain of $N$ nodes, as shown in \FIG\ref{fig:schem dim1}(a).
There is a link from node $i$ to node $i+1$ 
with the weight $w_{i,i+1}>0$ for each $1\le i\le N-1$.
There is a link from node $i$ to node $i-1$
with the weight $w_{i,i-i} >0$ for each
$2\le i\le N$. Generally, $w_{i,i+1}$ is not equal to $w_{i+1,i}$.
There are no other links.
There is only one spanning tree rooted at each node $i$, which is
represented by $1\gets 2\gets$ $\cdots\gets i-1\gets$ $i\to
i+1\to$ $\cdots \to N-1\to N$, where the arrow denotes either a
directed link or a directed path without confusion.  Therefore,
\begin{equation}
v_i= \frac{w_{2,1}w_{3,2}\cdots,
w_{i,i-1}w_{i,i+1}w_{i+1,i+2}\cdots
w_{N-1,N}}{{\cal N}},
\end{equation}
where the normalization constant is given by
\begin{equation}
{\cal N}= \sum^N_{i=1}w_{2,1}w_{3,2}\cdots,
w_{i,i-1}w_{i,i+1}w_{i+1,i+2}\cdots w_{N-1,N}.
\end{equation}
Note that the position of the node, whether
located in the middle or the periphery
of the chain, does not affect the value of $v_i$.

Consider a special case where
$w_{1,2}=w_{2,3}=\cdots=w_{N-1,N}=1$ and
$w_{2,1}=w_{3,2}=\cdots=w_{N,N-1}=\epsilon$ [\FIG\ref{fig:schem dim1}(b)].
For this network, we obtain
\begin{equation}
v_i=\frac{\epsilon^{i-1}(1-\epsilon)}{(1-\epsilon^N)}.
\label{eq:v_i chain special}
\end{equation}
When $\epsilon$ is small, node $i$ having a small $i$ is more influential.
With the normalization constant neglected, the MA yields
$k_1^{\rm out}/k_1^{\rm in}= 1/\epsilon$,
$k_2^{\rm out}/k_2^{\rm in}=\cdots=k_{N-1}^{\rm out}/k_{N-1}^{\rm in}
=1$, and $k_N^{\rm out}/k_N^{\rm in}=\epsilon$.
The MA is inconsistent with \EQ\eqref{eq:v_i chain special},
except under the limit $\epsilon\to 0$, in which case
$v_1\approx 1$, $v_2,\ldots, v_N\approx 0$.

\subsection{Weighted cycle}\label{sub:cycle}

Consider a weighted cycle having $N$ nodes, as depicted in
\FIG\ref{fig:schem dim1}(c). The weighted cycle is constructed by
adding two links $N\to 1$ and $1\to N$
with the weights $w_{N,1}$ and $w_{1,N}$, respectively, to
the weighted chain.

In this network, 
there are $N$ spanning trees rooted at node $i$, i.e.,
$j\gets j+1\gets \cdots i-1\gets i \to
i+1\to\cdots\to j-1$, where $1\le j\le N$;
nodes $N+1$ and 0 are identified with nodes 1 and $N$, respectively.
Therefore, we obtain
\begin{eqnarray}
v_i &\propto& w_{i,i+1}w_{i+1,i+2}\cdots w_{i-2,i-1}
+ w_{i,i-1}w_{i,i+1}\cdots w_{i-3,i-2}\nonumber\\
&+& w_{i-1,i-2}w_{i,i-1}w_{i,i+1}\cdots w_{i-4,i-3}+ \cdots,
\end{eqnarray}
where $w_{N,N+1}\equiv w_{N,1}$ and $w_{1,0}\equiv w_{1,N}$.

In the weighted chain, only the weights of the
descending links, i.e., $w_{j,j+1}$ for $j\ge i$ and
$w_{j+1,j}$ for $j+1\le i$, contribute to $v_i$. In contrast,
in the weighted cycle,
both $w_{j,j+1}$ and $w_{j+1,j}$ ($j,j+1\neq i$) 
contribute to $v_i$. Therefore, in the weighted cycle,
the effect of each link weight on $v_i$ is more blurred
than that in the case of the
weighted chain. This property comes from the fact that
node $i$ and node $j$ ($i\neq j$) are connected in two ways, i.e.,
clockwise and anticlockwise.

As a special case, consider a directed cycle in which
$w_{2,1}=w_{3,2}=\cdots=w_{N,N-1}=w_{1,N}=0$,
$w_{1,2}=w_{2,3}=\cdots=w_{N-1,N}=1$, and $w_{N,1}=\epsilon$
[\FIG\ref{fig:schem dim1}(d)].
In this network, the values of the influence are equal to
\begin{eqnarray}
v_1 &=& \frac{1}{1+(N-1)\epsilon},\\
v_2 =\cdots=v_N &=& \frac{\epsilon}{1+(N-1)\epsilon}.
\label{eq:cycle ex}
\end{eqnarray}
The MA, which yields
$k_1^{\rm out}/k_1^{\rm in}=1/\epsilon$,
$k_2^{\rm out}/k_2^{\rm in}=\cdots=
k_{N-1}^{\rm out}/k_{N-1}^{\rm in}=1$, and
$k_N^{\rm out}/k_N^{\rm in}=\epsilon$, is inconsistent with
\EQ\eqref{eq:cycle ex} except when $\epsilon\to 0$.

\subsection{Directed multipartite network}

\begin{figure}
\begin{center}
\includegraphics[width=6cm]{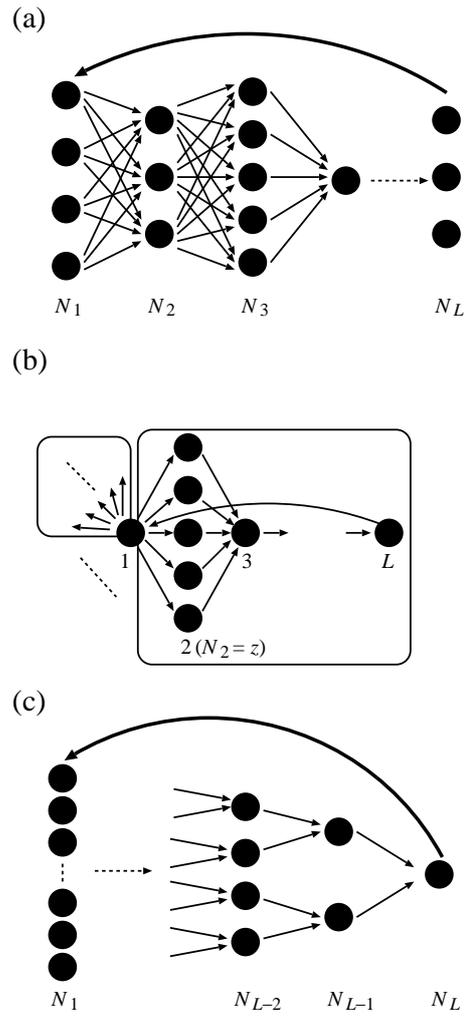}
\caption{Schematic of (a) multipartite network, (b) super-star,
and (c) funnel.}
\label{fig:schem multipart}
\end{center}
\end{figure}

Consider the directed $L$-partite network, as schematically shown in
\FIG\ref{fig:schem multipart}(a).  Layer $\ell$ ($1\le \ell\le L$) contains
$N_{\ell}$ nodes. Each node in layer $\ell$ sends directed links to all the
$N_{\ell+1}$ nodes in layer $\ell+1$, where layer $L+1$ is identified
as layer 1. Because of symmetry,
all nodes in layer $\ell$ have the same value of
influence, denoted by $v_{\ell}$. From \EQ\eqref{eq:influence},
we obtain
\begin{equation}
N_{\ell-1} v_\ell = N_{\ell+1} v_{\ell+1},\quad (1\le \ell\le L),
\label{eq:multi 1}
\end{equation}
where $N_0\equiv N_L$.
By combining \EQ\eqref{eq:multi 1} with the normalization condition
$\sum^L_{\ell=1}N_{\ell}v_{\ell}=1$, we obtain
\begin{equation}
v_{\ell}=\frac{1}{N_{\ell-1}N_{\ell}\sum^L_{\ell^{\prime}=1}
N_{\ell^{\prime}}^{-1}}.
\label{eq:v_i multipartite}
\end{equation}

\subsubsection{Super-star}

The super-star, which was introduced in \cite{Lieberman05} to study the
fixation probability in networks, is a variant of the
directed multipartite
network.  The super-star shown in \FIG\ref{fig:schem multipart}(b) is
generated as a superposition of a certain number of
identical directed multipartite
networks with $N_1=N_3=N_4=\cdots=N_L=1$ and $N_2=z$ ($\ge 1$). Each
multipartite network is called a
leave. The leaves are superposed such
that they share a
single node in layer 1. The indegree and outdegree of this node
are equal to the number of leaves.

It can be easily shown that
$v_i$ is independent of the number of leaves. Therefore, we consider
the case of a single leave.  Then, \EQ\eqref{eq:v_i multipartite}
yields
\begin{eqnarray}
v_1=v_4=v_5=\cdots=v_L &=& \frac{z}{z(L-1)+1},\\
v_2=v_3 &=& \frac{1}{z(L-1)+1}.
\end{eqnarray}
Surprisingly, the node in layer 1 does not have a particularly
large influence value. Given $z\ge 2$, the nodes in the expanded layer
(i.e., layer 2) and the node that receives convergent links from this
layer (i.e., layer 3) have small influence values. These relationships
are not predicted by the MA. The MA yields $k_2^{\rm out}/k_2^{\rm in}=
k_4^{\rm out}/k_4^{\rm in}=k_5^{\rm out}/k_5^{\rm in}=\cdots=k_N^{\rm out}/k_N^{\rm in}= 1$,
$k_1^{\rm out}/k_1^{\rm in}=z$, and $k_3^{\rm out}/k_3^{\rm in}=1/z$; 
the actual $v_1$
and $v_2$ values are essentially smaller than the values predicted by the
MA.

\subsubsection{Funnel}

The funnel, shown in \FIG\ref{fig:schem multipart}(c), was
introduced in \cite{Lieberman05} along with the
super-star; it is also a directed multipartite network.
The funnel has $N_{\ell}=z^{L-\ell}$ 
nodes in layer $\ell$ ($1\le \ell \le L$).
Using \EQ\eqref{eq:v_i multipartite}, we obtain
\begin{equation}
  v_{\ell} = 
  \begin{cases}
    \displaystyle{
      \frac{z - 1}{z^L - 1}}, \qquad
    & (\ell = 1), \\[3mm]
    \displaystyle{
      \frac{z - 1}{z^L - 1} \, z^{2\ell-L-2}}, \qquad
    & (2\le\ell\le L).
  \end{cases}
\end{equation}
The nodes in layer $L$ are most influential, and the node in layer
2 is least influential. The nodes in layer
1 are intermediately influential. For a large $z$,
they are as influential as a node in layer $\approx L/2$.
These relationships are not predicted by the MA,
which yields the following results:
$k_1^{\rm out}/k_1^{\rm in}=1$,
$k_2^{\rm out}/k_2^{\rm in}=\cdots=k_{L-1}^{\rm out}/k_{L-1}^{\rm in}=z^{-1}$,
and $k_L^{\rm out}/k_L^{\rm in}=z^{L-2}$.

\subsection{Directionally biased random network}\label{sub:korinet}

The networks considered in the previous sections have inherent global
directionality due to the presence of asymmetrically weighted or
unidirectional links from node to node or from layer to layer. The
directionality of networks is a main cause for the deviation in the
$v_i$ values from
those predicted by the MA.  To examine the effect of
directionality in further detail, we study the directionally
biased random network \cite{Kori06pre}. To generate a network from
this model, we prepare a strongly connected directed random graph with
mean indegree and mean outdegree $z$ and specify a root node, which is
placed in layer 1.  The root node is the source of directed links to about
$z$ nodes, which are placed in layer 2. We align all the nodes according
to their distance from the root. Except in the layers
near the last layer, the number
of nodes in layer $\ell$ grows roughly as $z^{\ell-1}$. We set the
weights of the forward links, i.e., links from layer $\ell$ to
layer $\ell+1$, as unity.  We set the weights of the backward links, i.e.,
links from layer $\ell$ to layer $\ell^{\prime}$, where
$\ell>\ell^{\prime}$, as $\epsilon$.  The weights of the parallel
links, i.e., those connecting two nodes in the same layer, are
arbitrary; they do not affect the value of $v_i$ in the
following derivation.
When $\epsilon=1$, the network is an unweighted directed
random graph, if the weight of the parallel link is equal to unity at
$\epsilon=1$.  When $\epsilon=0$, the network is purely feedforward
and no longer strongly connected. The feedforwardness is parametrized by
$\epsilon$.

\begin{figure}
\begin{center}
\includegraphics[width=6cm]{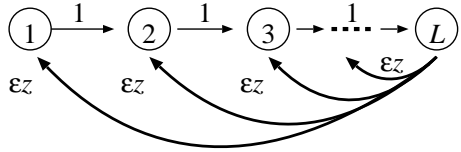}
\caption{Schematic of directionally biased
random network under tree approximation.}
\label{fig:1d-tree-a}
\end{center}
\end{figure}

The directionally biased random network is approximated by using a modified
tree as follows \cite{Kori06pre}. We assume that each node has $z$
outgoing links and that each node except the root node has only one
``parent'' node, namely, the node in the previous layer from where it
receives a feedforward link.  Further assume that there are
$L$ layers and that layer $\ell$ ($1\le \ell\le L$) has $z^{\ell-1}$
nodes. The number of nodes is equal to
$(z^L-1)/(z-1)$.  At this point, the constructed
network is a tree.  Then, we add backward links with weight $\epsilon$
to this tree. When $z$ is large, most backward links in the original
network originate from layer $L$, because layer $L$ has a majority of
nodes.  Therefore, we assume that, in the approximated network, 
the backward links with weight $\epsilon$ originate
only from the nodes in layer $L$.
This approximation is accurate when $z$ is sufficiently
large. The other links in the approximated network
have the weight of unity. For a sufficiently large $z$, all
nodes in the same
layer have almost the same connectivity pattern. In terms of incoming
links, a node receives approximately one forward link from the
previous layer and $z$ backward links from layer $L$.  The
approximated network is
schematically shown in \FIG\ref{fig:1d-tree-a}. On an average, each
node in layer $L$ is the source of an directed 
link to each node with an effective weight
$\epsilon^{\prime}$.  Because
$k_i^{\rm out}=\epsilon z$ for a node in layer $L$ is approximated by
$\epsilon^{\prime}(z^L-1)/(z-1)\approx
\epsilon^{\prime}z^{L-1}$, we obtain $\epsilon^{\prime}\approx
\epsilon z^{-L+2}$.
 
The influence of a node in layer $\ell$, denoted by $v_{\ell}$, satisfies
the following relationships:
\begin{eqnarray}
z^{L-1}\epsilon^{\prime} v_1 &=& z v_2,\\
(z^{L-1}\epsilon^{\prime} +1)v_{\ell} &=& 
z v_{\ell+1},\quad (2\le \ell\le L-1).
\label{eq:DBRN 1}
\end{eqnarray}
On substituting $\epsilon^{\prime}\approx\epsilon z^{-L+2}$ in
\EQ\eqref{eq:DBRN 1}
and considering the normalization given by
\begin{equation}
\sum_{\ell=1}^L z^{\ell-1} v_{\ell} = 1,
\end{equation}
we obtain
\begin{equation}
v_{\ell} \approx
  \begin{cases}
    \displaystyle{
      \left(\epsilon z +1\right)^{-L+1}}, \qquad
    & (\ell = 1), \\[3mm]
    \displaystyle{
      \left(\epsilon z+1\right)^{-L+\ell-1}
      z^{-\ell+2}\epsilon}, \qquad
    & (2\le\ell\le L).
  \end{cases}
\label{eq:v_i korinet}
\end{equation}

For a small $\epsilon$, the network is close to feedforward, and
$v_1$ is relatively large;
$v_{\ell}$ varies as
$v_{\ell}\propto \left(\epsilon+z^{-1}\right)^{\ell}$.
When $z$ is sufficiently large, we obtain
$v_{\ell}\propto \epsilon^{\ell}$,
which coincides with the results obtained for the
network shown in \FIG\ref{fig:schem dim1}(b) (\SEC\ref{sub:chain}). 

\begin{figure}
\begin{center}
\includegraphics[width=6cm]{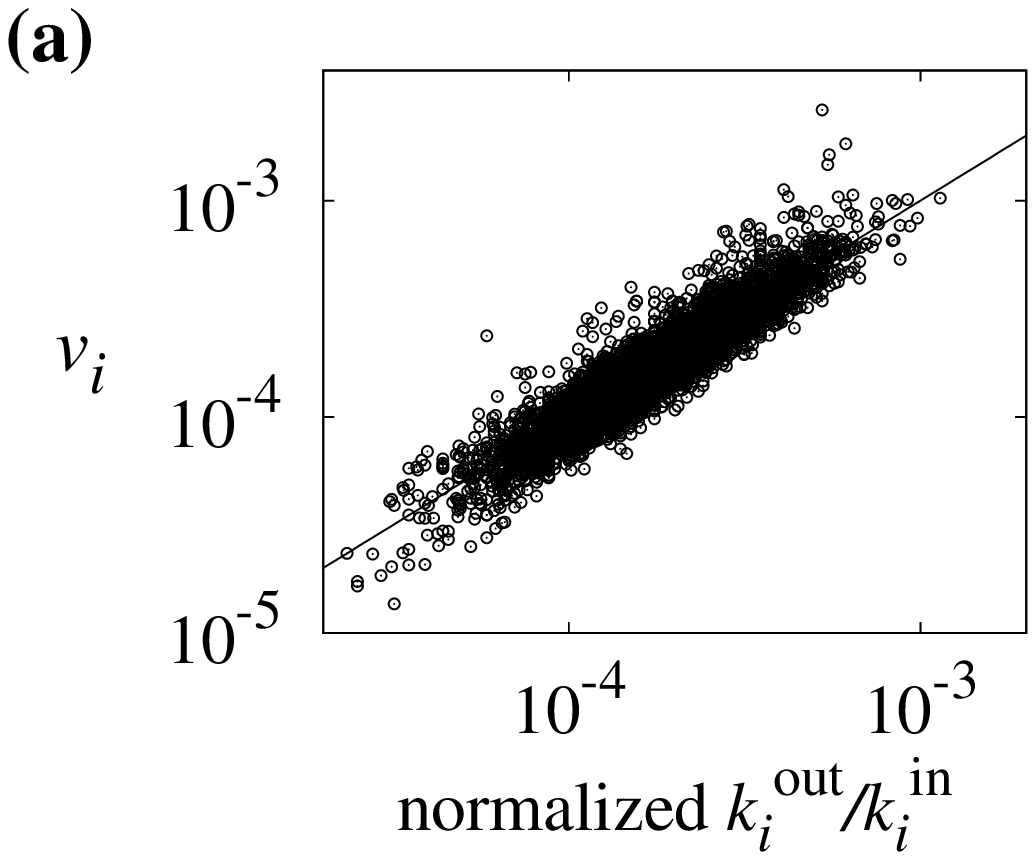}
\includegraphics[width=6cm]{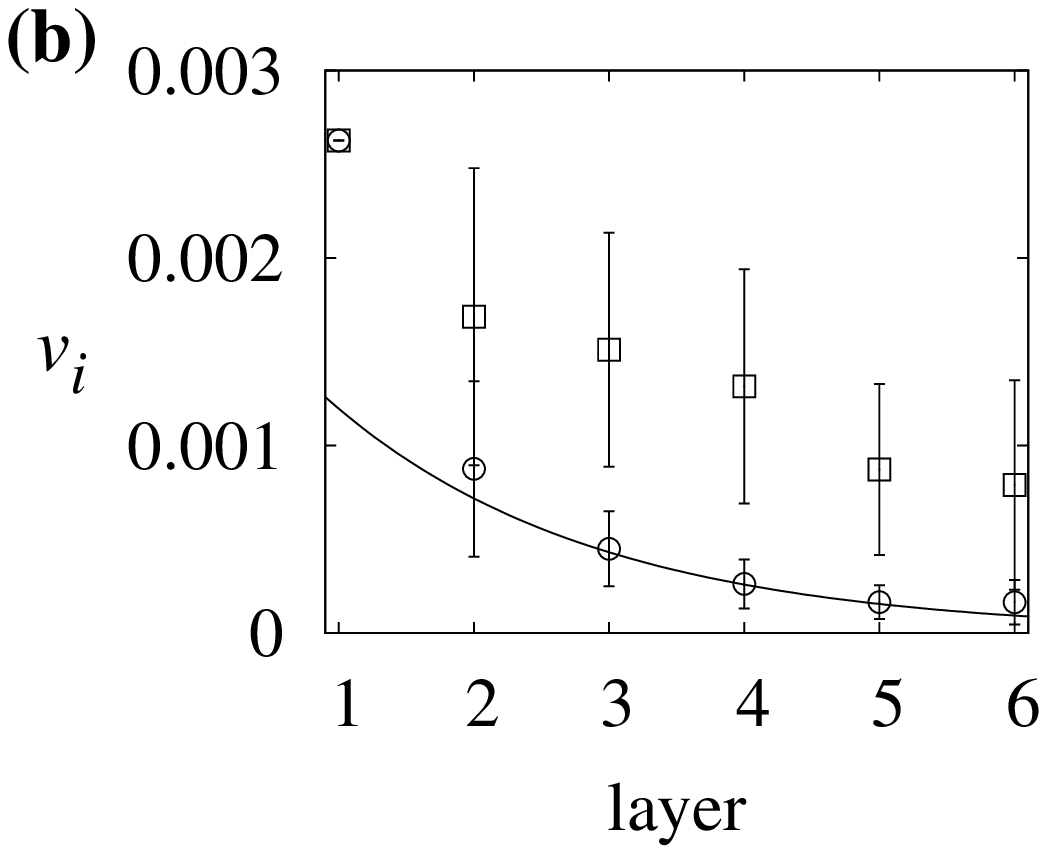}
\includegraphics[width=6cm]{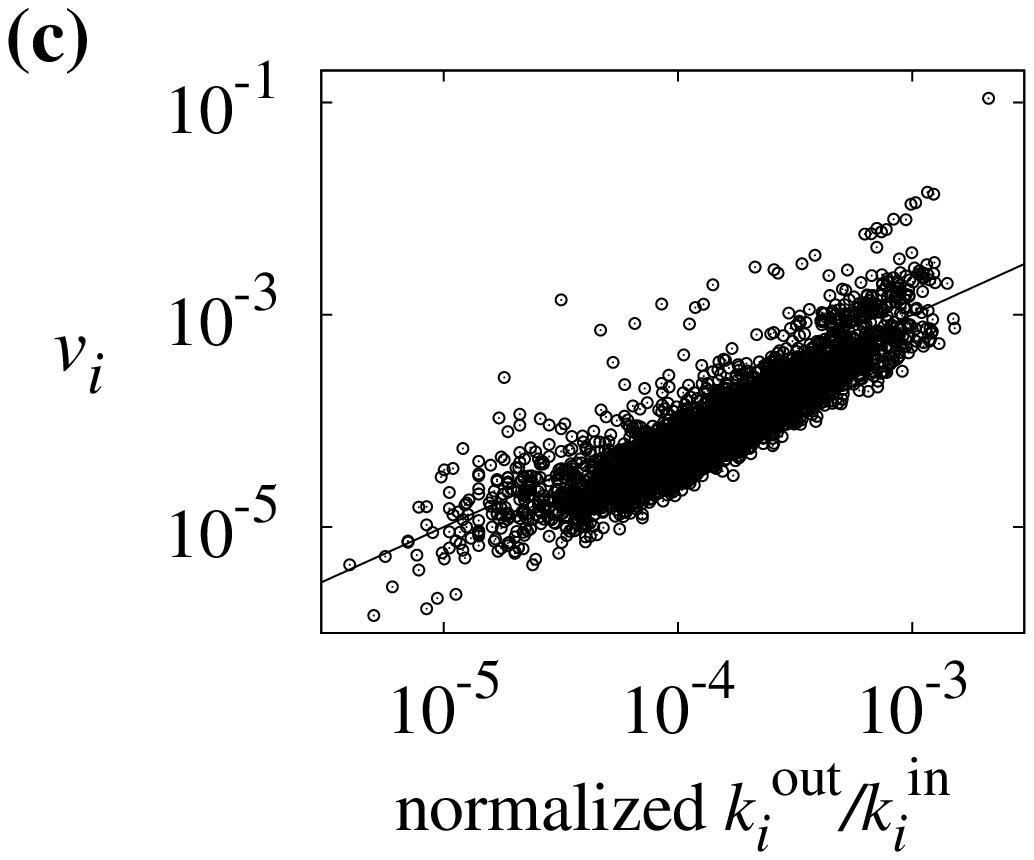}
\includegraphics[width=6cm]{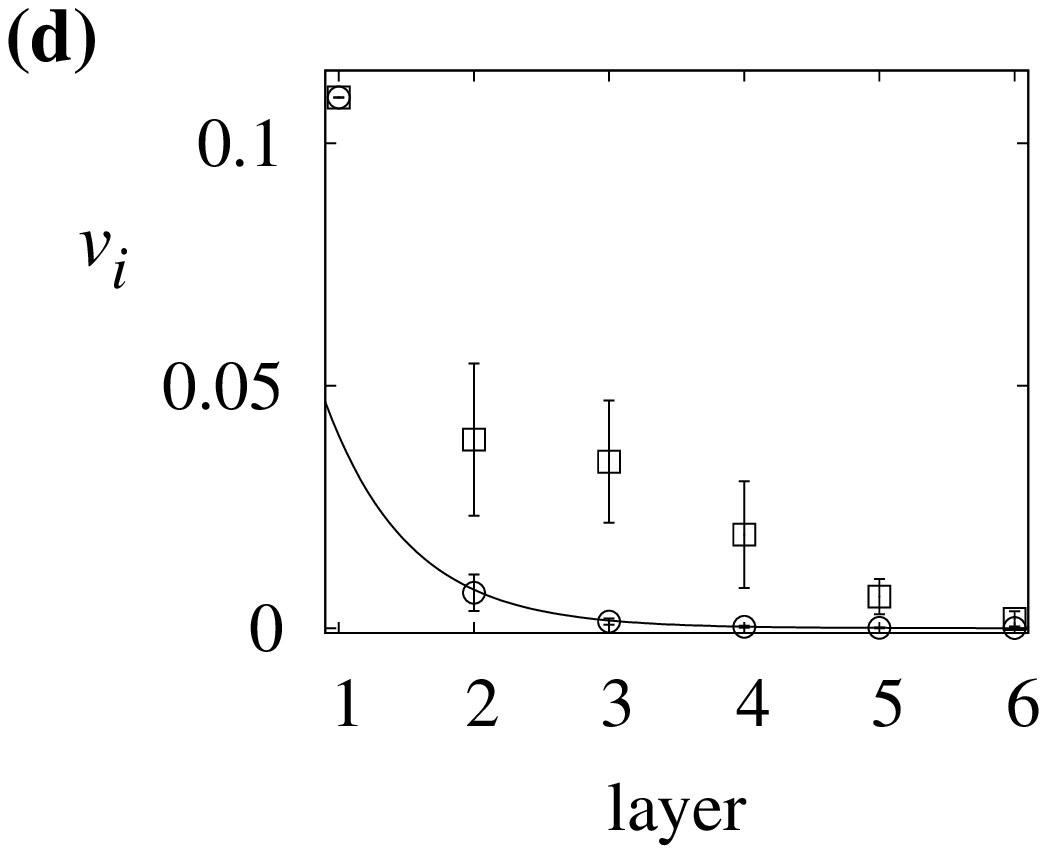}
\caption{$v_i$ for a directionally biased random network with
$N=5000$. We set [(a) and (b)] $\epsilon=0.5$ and [(c) and (d)]
$\epsilon=0.1$. In
(a) and (c), $v_i$ is plotted against the MA results.  The lines
represent the MA: $v_i=(k_i^{\rm out}/k_i^{\rm in}) /
\sum_{j=1}^N(k_j^{\rm out}/k_j^{\rm in})$.  In (b) and (d), $v_i$
(circles) and rescaled $k_i^{\rm out}/k_i^{\rm in}$ (squares) averaged
over all the nodes in each layer are plotted against layer number. The
error bars indicate the standard deviation obtained from $v_i$ of all
the nodes in the same layer.}
\label{fig:directional}
\end{center}
\end{figure}

To test our theory, we generate a directionally biased random network
with $N=5000$ and $z=10$.  For $\epsilon=0.5$, the values
of $v_i$ of all the nodes are plotted against the values obtained from
the MA in \FIG\ref{fig:directional}(a). 
Although the values obtained from the
MA are strongly
correlated with $v_i$, there is some variation in $v_i$ for a fixed
$k_i^{\rm out}/k_i^{\rm in}$.  The average and the standard deviation of
$v_i$ in each layer are plotted by the circles and the corresponding
error bars, respectively, in
\FIG\ref{fig:directional}(b).
The influence of a node decreases exponentially
with $\ell$ as predicted by \EQ\eqref{eq:v_i korinet}
[\EQ\eqref{eq:v_i korinet} for $\ell\ge 2$ is represented by the line in
\FIG\ref{fig:directional}(b)].  The average and the standard deviation
of $v_i$ obtained by 
the MA are plotted by the squares and the corresponding error bars,
respectively.
The values obtained from the MA are scaled by a
multiplicative factor $C$, where $C$ is selected such that
$v_i=Ck_i^{\rm out}/k_i^{\rm in}$ for the root node (i.e., $\ell=1$).
Figure~\ref{fig:directional}(b) shows that
$k_i^{\rm out}/k_i^{\rm in}$ is generally small for node $i$ 
in a downstream
layer. However, the decrease in $v_i$ with $\ell$ is much more than 
that in $k_i^{\rm out}/k_i^{\rm in}$.
The hierarchical nature of the
network is revealed by the $v_i$ values and not satisfactorily by the
local degree.  The results for $\epsilon=0.1$ shown in
\FIGS\ref{fig:directional}(c) and 
\ref{fig:directional}(d) provide further evidence 
for our claim.

\subsection{Small-world networks}\label{sub:sw}

\begin{figure}
\begin{center}
\includegraphics[width=9cm]{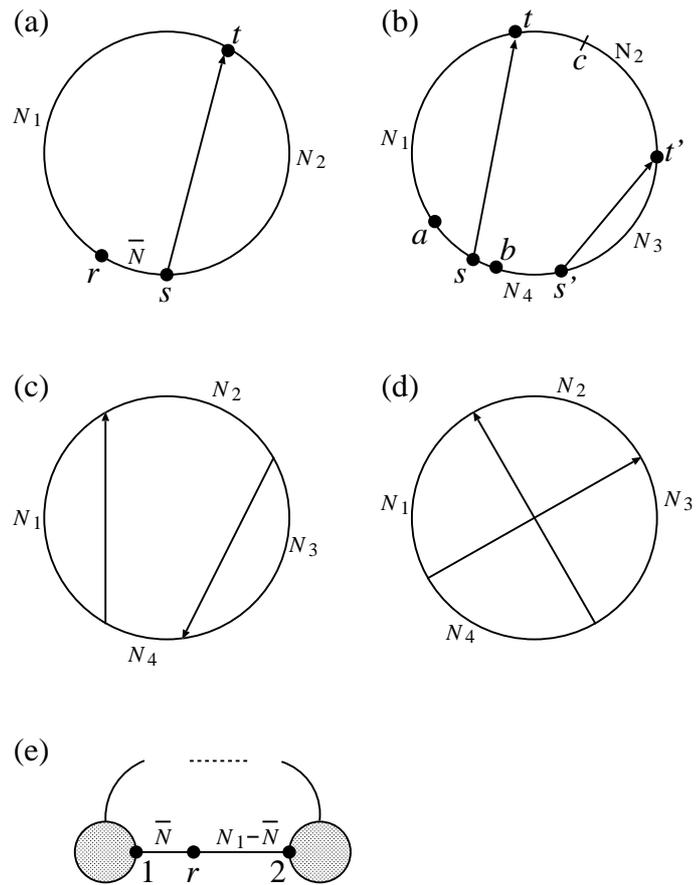}
\caption{Schematics of directed small-world network having (a) one
shortcut, [(b)--(d)] two shortcuts, and (e) general number of
shortcuts.}
\label{fig:schem sw}
\end{center}
\end{figure}

In this section,
we analyze the influence in the directed unweighted 
small-world network model,
which is a variant of the Watts-Strogatz model \cite{Watts98}.
To generate a network, we start with an undirected cycle of $N$
nodes, in which each node is connected to its immediate neighbor on
both sides. At this stage, $k_i^{\rm in}=k_i^{\rm out}=2$ is satisfied
for all $i$. Then,
we add a directed shortcut to the network, as schematically shown in
\FIG\ref{fig:schem sw}(a).  The source and the target of the shortcut are
denoted by nodes $s$ and $t$, respectively. The distance between node $s$
and node $t$ along the cycle is assumed to be 
$\min(N_1,N_2)$, where
$N_2\equiv N-N_1$.

We enumerate the number of directed
spanning trees rooted at node $r$, 
which is $\overline{N}$ nodes away from
node $s$ along the cycle, where
$0\le \overline{N}\le \max(N_1,N_2)$.
There are $N$ spanning trees that do not use
the shortcut, as derived in \SEC\ref{sub:cycle}.  Any
spanning tree that uses the shortcut includes the directed path
$r\to \cdots \to s$, which contains $\overline{N}+1$ nodes.
The choice of the other links is
arbitrary with the restriction that a spanning tree must be formed.  The
$N_1-\overline{N}-1$ nodes between node $r$ and node $t$ in
\FIG\ref{fig:schem sw}(a) are reached from node $r$ or node $t$
by a directed path along the cycle. There are $N_1-\overline{N}$ choices
regarding the formation of this part of the spanning tree.
The $N_2-1$ nodes between node $s$ and node $t$
are reached from node $s$ or node $t$ by a directed path along the cycle.
There are $N_2$ choices regarding the formation of this part of the
spanning tree. In sum, there are $N+(N_1-\overline{N})N_2$ spanning
trees rooted at $v$. Therefore, the influence of $v$
is large (small)
for the $v$ that is close
to the source (target) of the shortcut. In the region on
the cycle where no source or target of a shortcut is located,
the influence of a node changes linearly with the distance between the 
source of the shortcut and the node, because
$N+(N_1-\overline{N})N_2\propto -\overline{N}$.
We call such a region, including the two border points, the segment.

Next, we consider small-world networks with two directed shortcuts. There
are three qualitatively different possible
arrangements of the shortcuts, as shown
in \FIGS\ref{fig:schem sw}(b)--\ref{fig:schem sw}(d).
The lengths of the four segments
are denoted by $N_1$, $N_2$, $N_3$, and
$N_4$, such that $N_1+N_2+N_3+N_4=N$.

In the network shown in \FIG\ref{fig:schem sw}(b), 
a node is located at either of the three essentially
different positions denoted by $a$, $b$, and $c$. 
We first enumerate spanning trees rooted at node $a$.
The distance from node $a$ to the source of 
a shortcut, i.e., node $s$, is denoted
by $\overline{N}$ ($0\le \overline{N}\le N_1$).
There are $N$ spanning trees that do not use the shortcuts.
There are $(N_1-\overline{N})(N_2+N_3+N_4)$ spanning trees 
that use the shortcut $s\to t$ but not $s^{\prime}\to
t^{\prime}$. There are
$(N_1+N_2-\overline{N})N_3$ spanning trees that use the shortcut
$s^{\prime}\to t^{\prime}$ but not $s\to t$.
There are $(N_1-\overline{N})N_2N_3$
spanning trees that use both shortcuts, which can be explained
as follows.
The directed path $a\to s\to s^{\prime}$ is included in
such a spanning tree. The $N_1-\overline{N}-1$ nodes between
node $a$ and node $t$ are reached along the cycle from 
node $a$ or node $t$. The $N_2-1$ nodes between
node $t$ and node $t^{\prime}$ are reached along the cycle from 
node $t$ or node $t^{\prime}$. The $N_3-1$ nodes between
node $t^{\prime}$ and node $s^{\prime}$ 
are reached along the cycle from node $t^{\prime}$ or 
node $s^{\prime}$.
In sum, the number of spanning trees rooted at node $a$ is equal to
\begin{equation}
N+ (N_1-\overline{N})(N_2+N_3+N_4) 
+ (N_1+N_2-\overline{N})N_3+(N_1-\overline{N})N_2N_3,
\label{eq:sw two shortcuts 1}
\end{equation}
which is proportional to the influence of node $a$.
If $N$ is sufficiently large and the two shortcuts are randomly placed,
the last term in \EQ\eqref{eq:sw two shortcuts 1}
is of the highest order
because $N_i=O(N)$ ($1\le i\le 4$). As in the case of the network
with one shortcut,
the influence changes linearly within one segment.

Similarly, the number of spanning trees rooted at node $b$ is equal to
\begin{equation}
N+ N_1(N_2+N_3+N_4-\overline{N}) 
+ (N_1+N_2+\overline{N})N_3+N_1N_2N_3,
\label{eq:sw two shortcuts 2}
\end{equation}
where $\overline{N}$ ($0\le \overline{N}\le N_4$)
is the distance from node $b$ to node $s$ along the cycle.
The number of spanning trees rooted at node $c$ is equal to
\begin{equation}
N+(N_2-\overline{N})N_3+N_1 \overline{N},
\label{eq:sw two shortcuts 3}
\end{equation}
where $\overline{N}$ ($0\le \overline{N}\le N_2$) 
is the distance from node $c$ to node $t$ along the cycle.
Owing to the absence of a third-order term in
\EQ\eqref{eq:sw two shortcuts 3},
the influence values of the nodes located on the segment between
the two targets of the shortcuts are very small.

The quantities given in 
\EQS\eqref{eq:sw two shortcuts 1}--\eqref{eq:sw two shortcuts 3}
are linear in $\overline{N}$. Therefore, the influence changes
linearly within each segment. This is true for the other two
types of arrangements of shortcuts
shown in \FIGS\ref{fig:schem sw}(c) and \ref{fig:schem sw}(d).

This linear relationship
also holds true in the case of more than two shortcuts. 
To show this, we consider a general directed small-world network and
focus on a segment on the cycle, which is
schematically shown in \FIG\ref{fig:schem sw}(e).
Without loss of generality, we assume that
a node $v$ in the segment is located
$\overline{N}$ and $N_1-\overline{N}$ nodes away from the two
border points of the segment.
We distinguish three types of spanning trees rooted at node $r$.
First, some spanning trees include both $r\to\cdots\to 1$ and
$r\to\cdots\to 2$. We denote
the number of such spanning trees
by $S_1$. Second, some spanning trees include
$r\to\cdots\to 1$ and not $r\to\cdots\to 2$.
For these spanning trees, node $2$ is reached via a path
$r\to 1\to\cdots\to 2$. To enumerate such spanning trees, 
denote by $S_2$ the number of directed
trees that span the network excluding the nodes in
the segment between
node $r$ and node $2$. Third, the other spanning trees include
$r\to\cdots\to 2$ and not
$r\to\cdots\to 1$. Denote by $S_3$
the number of directed trees that span the network excluding the nodes
in the segment between node $r$ and node $1$. 
The number of spanning trees rooted at node $r$ is equal to
\[ S_1+S_2(N_1-\overline{N})+S_3\overline{N}.\]
Therefore, the influence of node $r$ changes linearly with
$\overline{N}$ ($0\le \overline{N}\le N_1$).

\begin{figure}
\begin{center}
\includegraphics[width=6cm]{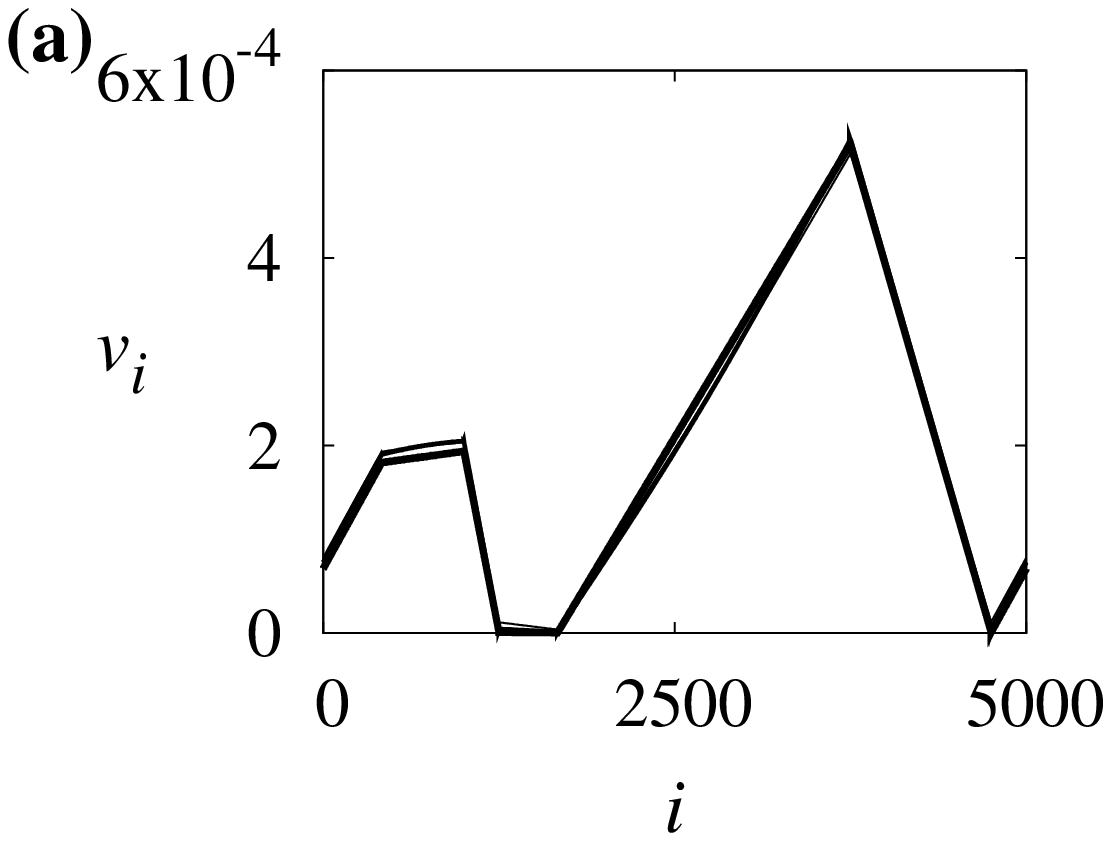}
\includegraphics[width=6cm]{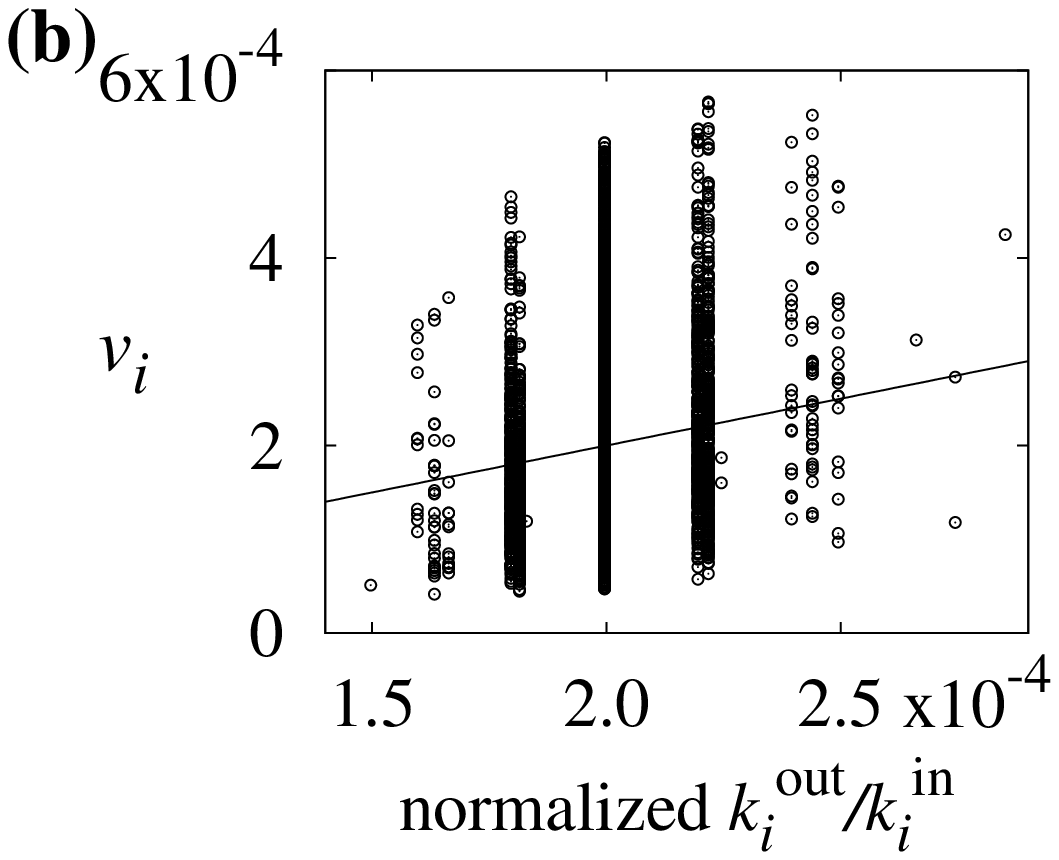}
\includegraphics[width=6cm]{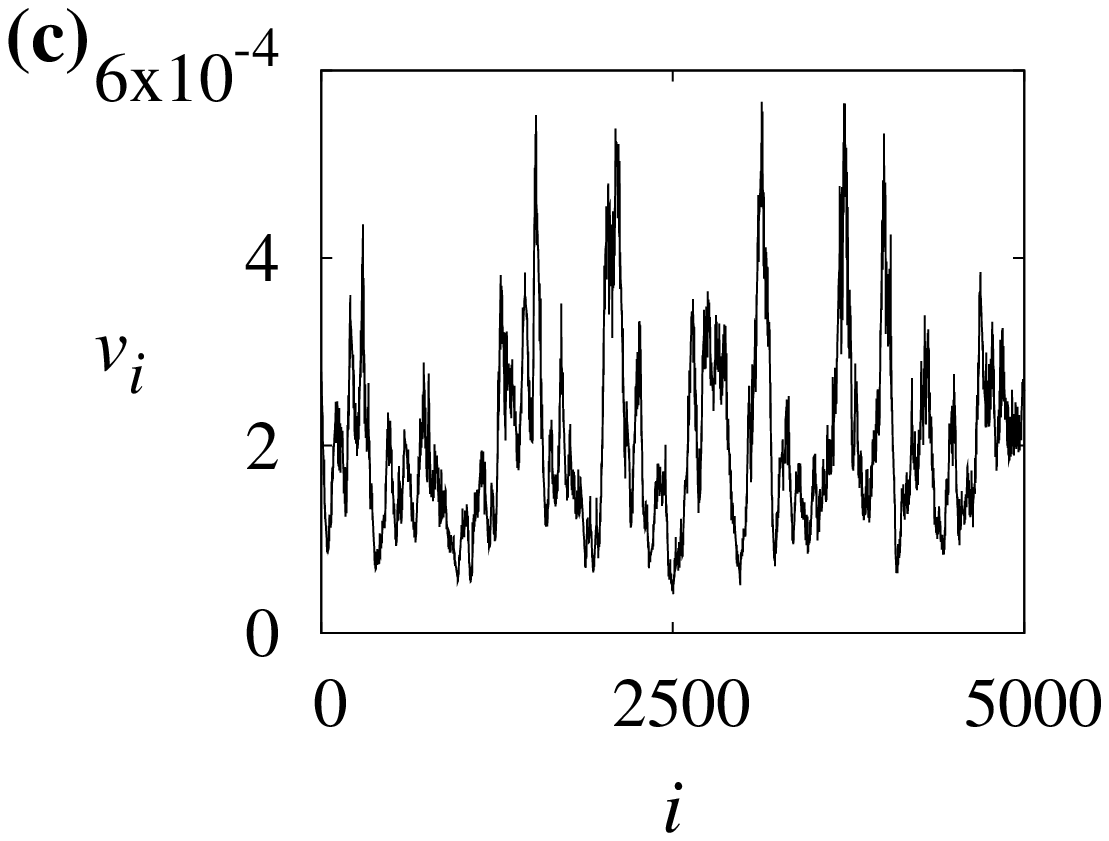}
\includegraphics[width=6cm]{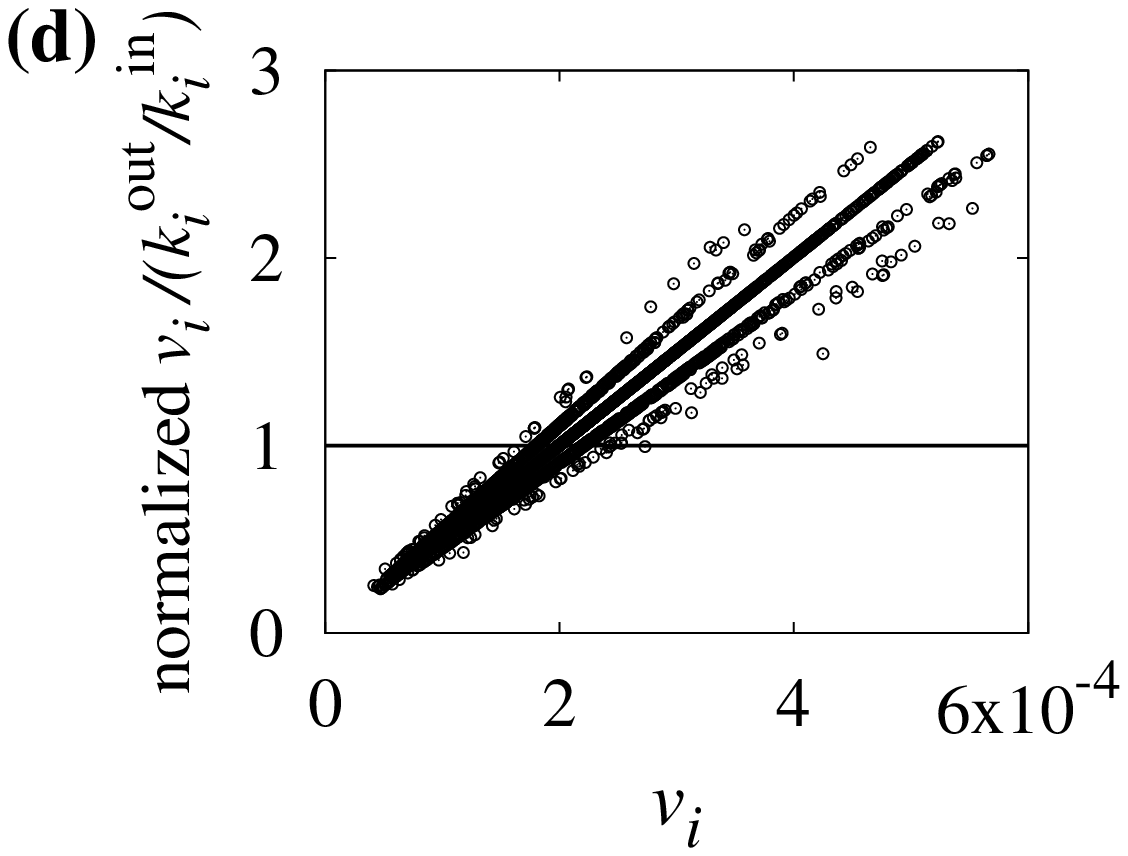}
\caption{$v_i$ in directed small-world networks with $N=5000$.  (a)
Results for small-world network with three added shortcuts. The thick,
medium, and thin lines correspond to the networks in which the mean
degree of the underlying cycle is equal to 2, 4, and 6,
respectively. (b, c, d) Results for directed small-world network with
$\left<k\right>=10$ and 500 rewired shortcuts. In (a) and (c), $v_i$
is plotted against the position of the node in the underlying
cycle. In (b), $v_i$ is plotted against $(k_i^{\rm out}/k_i^{\rm in})
/ \sum_{j=1}^N(k_j^{\rm out}/k_j^{\rm in})$; the line represents the
MA result. In (d), the relationship between $v_i/(k_i^{\rm
out}/k_i^{\rm in})$ and $v_i$ is shown.}
\label{fig:sw}
\end{center}
\end{figure}

The thick line in \FIG\ref{fig:sw}(a) indicates
the numerically obtained
values of $v_i$ for a small-world network
with three shortcuts. We set $N=5000$.
The nodes are aligned according to their position in the cycle.
In accordance with the theoretical prediction,
$v_i$ changes linearly with $i$ within
each segment. 
$v_i$ is very small for $1250\le i\le 1666$, because these nodes are
between two targets of shortcuts.

In theory, it is assumed that each node is initially
connected to only its nearest neighbors on the cycle (i.e.,
$k_i^{\rm in}=k_i^{\rm out}=2$). However, values of $v_i$ are almost the
same if the underlying cycle has
$k_i^{\rm in}=k_i^{\rm out}=4$ (i.e., each node is connected to
two neighbors on each side) or 
$k_i^{\rm in}=k_i^{\rm out}=6$. The results for
$k_i^{\rm in}=k_i^{\rm out}=4$ and those for $k_i^{\rm in}=k_i^{\rm out}=6$
are indicated by the medium and thin lines, respectively, in 
\FIG\ref{fig:sw}(a). The three lines are observed to almost overlap with
each other. 

To examine the effect of shortcuts in more general small-world networks,
we generate a small-world network by
rewiring many links \cite{Watts98}. We place $N=5000$ nodes on a cycle and
connect a node to its five immediate neighbors on each side, such that
$k_i^{\rm in}=k_i^{\rm out}=10$. 
Then, out of 50000 directed links,
we rewire 500 randomly selected ones
to create directed shortcuts. The sources and targets of shortcuts
are chosen randomly from the network with the restriction that
self loops and multiple links must be avoided.
Because of rewiring, the mean degree
$\left<k\right>=10$.

As shown in \FIG\ref{fig:sw}(b), the MA strongly disagrees with the
observed $v_i$ \cite{Masuda09njp}.  The values of $v_i$ are
plotted against the circular positions of the nodes in
\FIG\ref{fig:sw}(c).  $v_i$ changes gradually along the cycle, which
is consistent with our analytical results. The peaks and troughs
in \FIG\ref{fig:sw}(c) correspond to the sources
and targets of the shortcuts, respectively.
For a node near a source (target), $v_i$ is
large (small), whereas the MA estimate $\propto k_i^{\rm out}/k_i^{\rm in}$ is
not as affected by the position of the shortcuts as $v_i$.
The relationship between
$v_i/(k_i^{\rm out}/k_i^{\rm in})$ and $v_i$ is shown in \FIG\ref{fig:sw}(d).
The MA is exact along the horizontal line, i.e.,
$v_i=(k_i^{\rm out}/k_i^{\rm in})/(\sum_{j=1}^N k_j^{\rm out}/k_j^{\rm in})$. 
Nodes with large (small) $v_i$ tend
to be located near sources (targets) of shortcuts.  For such nodes,
$v_i$ is usually larger (smaller) than the value obtained by the MA.

\section{Entrainment of a network by a pacemaker}\label{sec:kcr}

As an application of the influence as a centrality measure,
we examine a system of
coupled phase oscillators having a pacemaker \cite{Kori06pre,Kori04prl}.
Consider a dynamical system of phase oscillators given by
\begin{equation}
  \dot{\phi}_i = \omega_i + \frac{\kappa}{\left<k\right>}
  \sum_{j=1}^N w_{ji} 
\sin\left( \phi_j - \phi_i \right), \qquad
(1\le i\le N),
\label{eq:coupled oscillators}
\end{equation}
where the mean degree
$\left<k\right>=\sum_{i^{\prime},j^{\prime}=1}^N
w_{i^{\prime}j^{\prime}}
/N$ provides the normalization for the coupling strength $\kappa$.
The phase and the intrinsic frequency of the $i$th oscillator are
denoted by $\phi_i\in
[0,2\pi)$ and $\omega_i$, respectively.
We assume a pacemaker, i.e., an oscillator
that is not influenced by the other oscillators, in the network.
Equation~\eqref{eq:coupled oscillators} emulates a pacemaker system,
where the pacemaker is placed at node $i_0$, if we force $w_{j i_0}=0$
($1\le j\le N$).  We examine the possibility
of the pacemaker to entrain the other oscillators into its own intrinsic
rhythm. For this, we assume that $\omega_i=\omega$
($i\neq i_0$) is identical for the $N-1$ oscillators and
that $\omega_{i_0}$ takes a different value.
By redefining $\phi_i-\omega t$ as the new $\phi_i$ and
rescaling time, we set
$\omega_{i_0}=1$ and $\omega_i=0$ ($i\neq i_0$) without loss of
generality.

Depending on the network and the position of the pacemaker,
there exists a critical threshold $\kappa_{\rm cr}$
such that the entrainment is realized for $\kappa>\kappa_{\rm cr}$
\cite{Kori04prl,Kori06pre}.
When entrained, the actual frequency of all
oscillators becomes exactly the same as that of the pacemaker,
i.e., $\omega_{i_0}=1$. Thus,
the condition for the entrainment is given by
\begin{equation}
\dot{\phi}_i = \frac{\kappa}{\left<k\right>} \sum_{j=1}^N w_{ji} 
\sin\left( \phi_j - \phi_i \right)=1, \qquad
(1\le i\le N, i\neq i_0).
\label{eq:entrainment}
\end{equation}

In general, 
entrainment \cite{Kori06pre,Masuda07jcn} and synchrony
\cite{syn-forward-pre} are easily
realized for feedforward networks.
Because of the intuitive meaning of the influence,
$\kappa_{\rm cr}$ may be small if $v_{i_0}$ is large.  We
analytically show this for the directionally biased random network
with a sufficiently large $z$.
In the directionally biased random network,
a node in layer $\ell$ receives a forward link with weight unity
from layer $\ell-1$.
Although a forward link is absent for a node in layer 1, this factor
is negligible
because layer 1 contains only one node. 
Most backward links with weight $\epsilon$ to a node in 
layer $\ell$, where $\ell\le L-1$,
originate from layer $L$, as discussed in 
\SEC\ref{sub:korinet}. The number of parallel links is smaller
than that of backward links.
We assume that the weight of the parallel link, which was assumed
to be arbitrary in \SEC\ref{sub:korinet}, is equal to $\epsilon$,
such that all the incoming links to 
a node in layer $L$, except one forward link, also have weight
$\epsilon$. Under this condition,
we approximate
$\left<k\right>\approx 1+\epsilon z$.

Denote by $\kappa_{\rm cr}^{(\ell)}$ the typical critical coupling
strength when the pacemaker is located at a node in layer $\ell$ in a
directionally biased random network.
First, we consider the case in which node $i_0$ coincides 
with the root node in the directionally biased random network, i.e.,
$\ell=1$. This case was analyzed in our previous studies
\cite{Kori06pre,Kori04prl}.
In the entrained state, the phase difference between 
the oscillators in the same layer is small, and the phases of the 
oscillators in layers with small $\ell$ are more advanced.
Therefore, we assume that
the phases of all the oscillators in the same layer are identical.
We set the difference between the phase of the
oscillator in layer $\ell$ and that in layer $\ell+1$ to
$\varDelta \phi_{\ell}$. The entrainment occurs if and only if
$\varDelta \phi_1$, $\ldots$,
$\varDelta \phi_{L-1}$ stay constant in a long run.
We obtain \cite{Kori06pre,Kori04prl}
\begin{equation}
  \varDelta \phi_{\ell} = \frac{
\left( 1 + \epsilon z \right)^{L-\ell}}{\kappa}, \qquad
(1\le \ell\le L-1).
\label{eq:1 pre}
\end{equation}
By applying the threshold condition $\varDelta \phi_1=1$, we obtain
\begin{equation}
  \kappa_{\rm cr}^{(1)} = \left( 1 + \epsilon z \right)^{L-1}.
  \label{eq:1}
\end{equation}

Next, we consider the case in which node $i_0$ is located in layer $L$. We
expect $\kappa_{\rm cr}^{(L)}$ to be large because the pacemaker is
located downstream of the network.  To analyze this
case, we redraw the network as a directionally biased
random network, such that node $i_0$ is located at the root. Then, 
statistically, the
network is the same as the original directionally biased
random network in terms of the positions of the nodes and the 
links. However, the effective link weight
in the redrawn network is equal to $\epsilon$,
because most links in the original network are backward links with
weight $\epsilon$.
By assuming that all links in the redrawn network have weight
$\epsilon$, the result for unweighted
directed random graph \cite{Kori06pre,Kori04prl} 
translates into
\begin{equation}
  \varDelta \phi_{\bar{\ell}} =
\frac{(1+\epsilon z)
\left( 1 + z \right)^{L-\bar{\ell}-1}}{\kappa\epsilon}, \qquad
(1\le \bar{\ell}\le L-1),
\label{eq:L pre}
\end{equation}
where $\bar{\ell}$ is 
the layer number in the redrawn network.
Analogous to the derivation of \EQ\eqref{eq:1} from \EQ\eqref{eq:1 pre},
from \EQ\eqref{eq:L pre}, we derive
\begin{equation}
  \kappa_{\rm cr}^{(L)} = \frac{(1+\epsilon z)
\left( 1+z \right)^{L-2}}{\epsilon}.
  \label{eq:L}
\end{equation}

\begin{figure}
\begin{center}
\includegraphics[width=6cm]{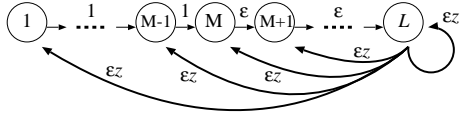}
\caption{Schematic of redrawn directionally biased random network when
the pacemaker is located in layer $L-M+1$ in the original
directionally biased random network.}
\label{fig:1d-tree-b}
\end{center}
\end{figure}

When node $i_0$ is located in the $(L-M+1)$th layer
$(2 \leq M \leq L-1)$ in
the original network, we redraw the network in a similar manner, such
that node $i_0$
is located at the root.
The redrawn network is schematically shown in
\FIG\ref{fig:1d-tree-b}. For this network, we obtain
\begin{equation}
  \varDelta \phi_{\bar{\ell}} = 
  \begin{cases}
\frac{\left( 1+z \right)^{L-M} 
\left( 1 + \epsilon z \right)^{M-\bar{\ell}}}{\kappa}, \qquad
    & (1\le\bar{\ell}\le M-1),\\
    \frac{(1+\epsilon z)\left( 1+z \right)^{L-\bar{\ell}-1}}{\kappa\epsilon}, \qquad
    & (M\le\bar{\ell}\le L-1),
  \end{cases}
\end{equation}
which yields
\begin{equation}
  \kappa_{\rm cr}^{(L-M+1)} = \left( 1 + \epsilon z \right)^{M-1}
\left( 1 + z \right)^{L-M}.
  \label{eq:L-M+1}
\end{equation}

From \EQS\eqref{eq:1}, \eqref{eq:L}, and
\eqref{eq:L-M+1}, we obtain
\begin{equation}
  \kappa_{\rm cr}^{(\ell)} = 
  \begin{cases}
    \left( 1 + \epsilon z \right)^{L-\ell}
\left( 1+z \right)^{\ell -1}, \qquad
    & (1\le\ell\le L-1),\\
    \frac{\left(1+\epsilon z\right) \left( 1+z \right)^{L-2}}{\epsilon}, \qquad
    & (\ell = L).
  \end{cases}
\label{eq:kcr summary}
\end{equation}
Under the condition
$\epsilon z \gg 1$,
\EQ\eqref{eq:kcr summary} gives
\begin{equation}
  \kappa_{\rm cr}^{(\ell)} \approx \epsilon^{L-\ell} \, z^{L-1}, \qquad
  (1\le\ell\le L).
\end{equation}
Moreover, \EQ\eqref{eq:v_i korinet} yields
\begin{equation}
  v_{\ell} \approx \epsilon^{-L+\ell} \, z^{-L+1}, \qquad
  (1\le\ell\le L).
\end{equation}
Therefore, we obtain
\begin{equation}
\kappa_{\rm cr}^{(\ell)} \approx \frac{1}{v_{\ell}}, \qquad
(1\le\ell\le L).
\label{eq:kcr vs v_i}
\end{equation}

Equation~\eqref{eq:kcr vs v_i} shows that the pacemaker located at an
influential node can easily realize the entrainment.  We validate this
prediction by direct numerical simulations of the pacemaker system on
a directionally biased random network with $N=200$, $z=10$, and
$\epsilon=0.1$. To judge whether the entrainment has been realized for a
value of $\kappa$, we measure the ratio of $\sum_{i=1,i\neq i_0}^N
\left[\phi_i\left(t=T\right)-\phi_i\left(t=0.8T\right)\right]\big/\left(N-1\right)$
to $\phi_{i_0}(t=T) - \phi_{i_0}(t=0.8 T)$, where $T$ is the duration
of a run.  The first 80\% of a run is discarded as transient. The
ratio represents the average phase shift of the oscillators, other than
the
pacemaker, relative to that of the pacemaker.  If this value is more
than 0.99, we consider the entrainment to be achieved.  Because the
transient is shorter for larger $\kappa$, we set $T=5\times
10^5/\kappa$.

\begin{figure}
\begin{center}
\includegraphics[width=6cm]{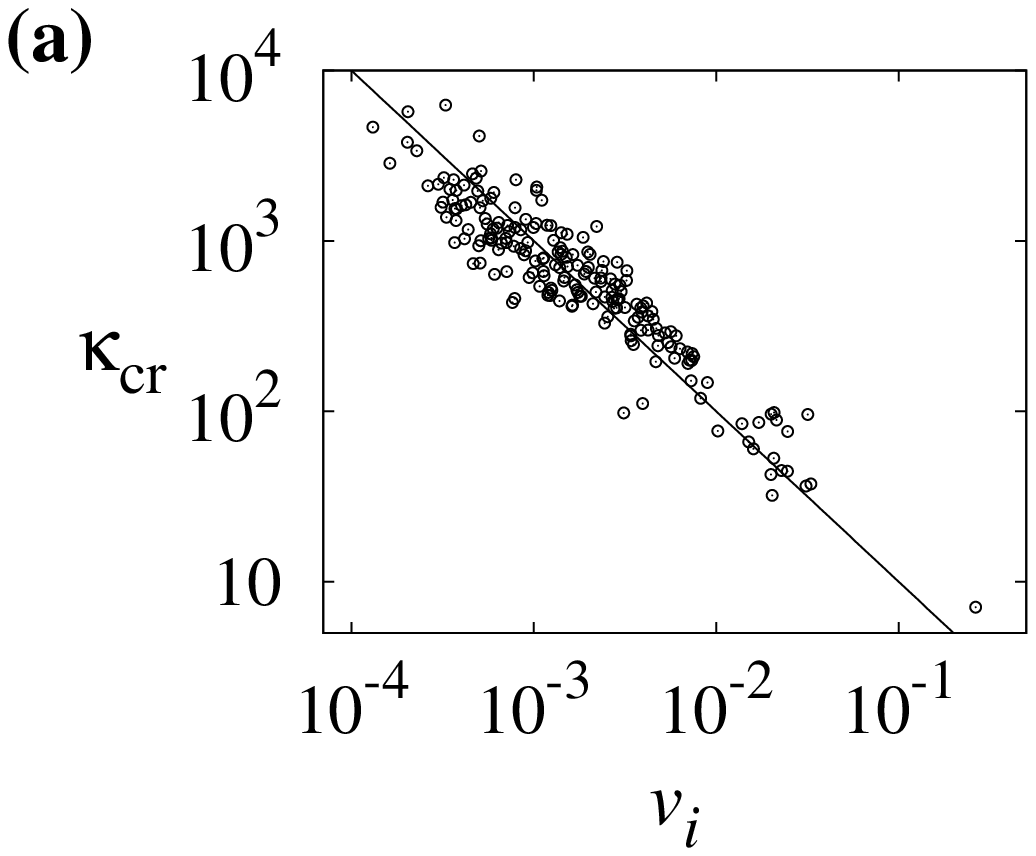}
\includegraphics[width=6cm]{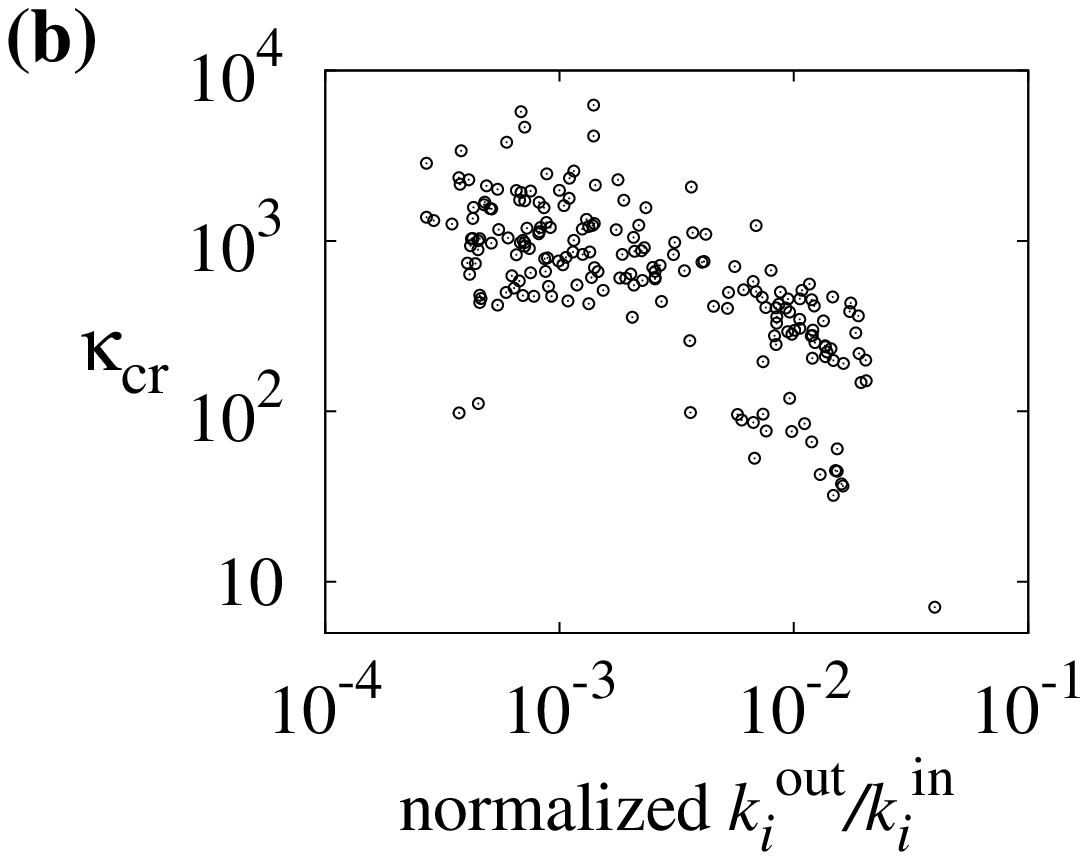}
\includegraphics[width=6cm]{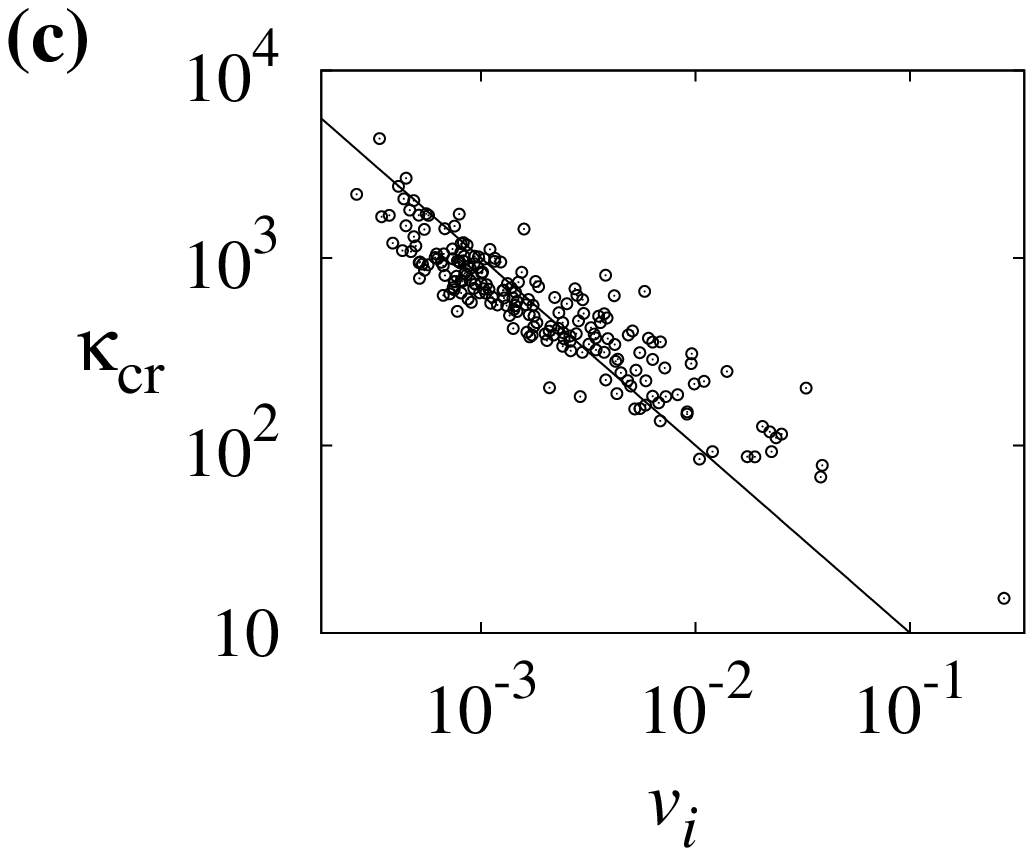}
\includegraphics[width=6cm]{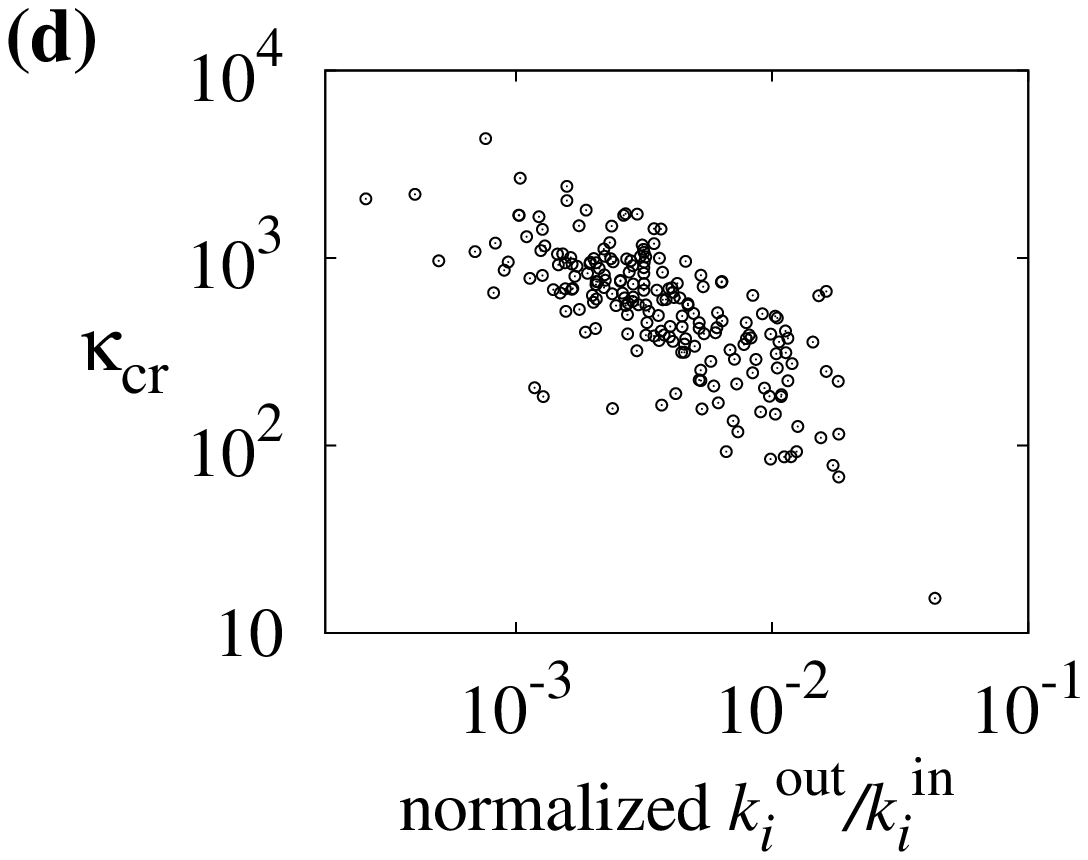}
\includegraphics[width=6cm]{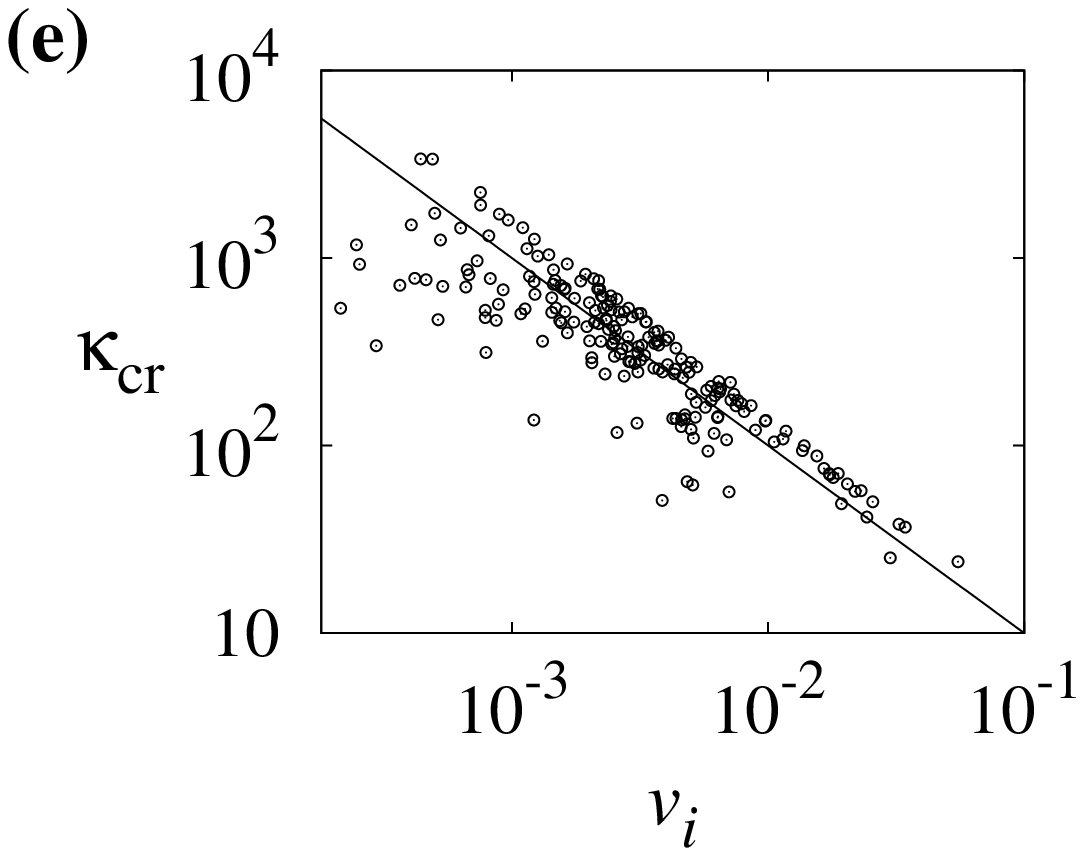}
\includegraphics[width=6cm]{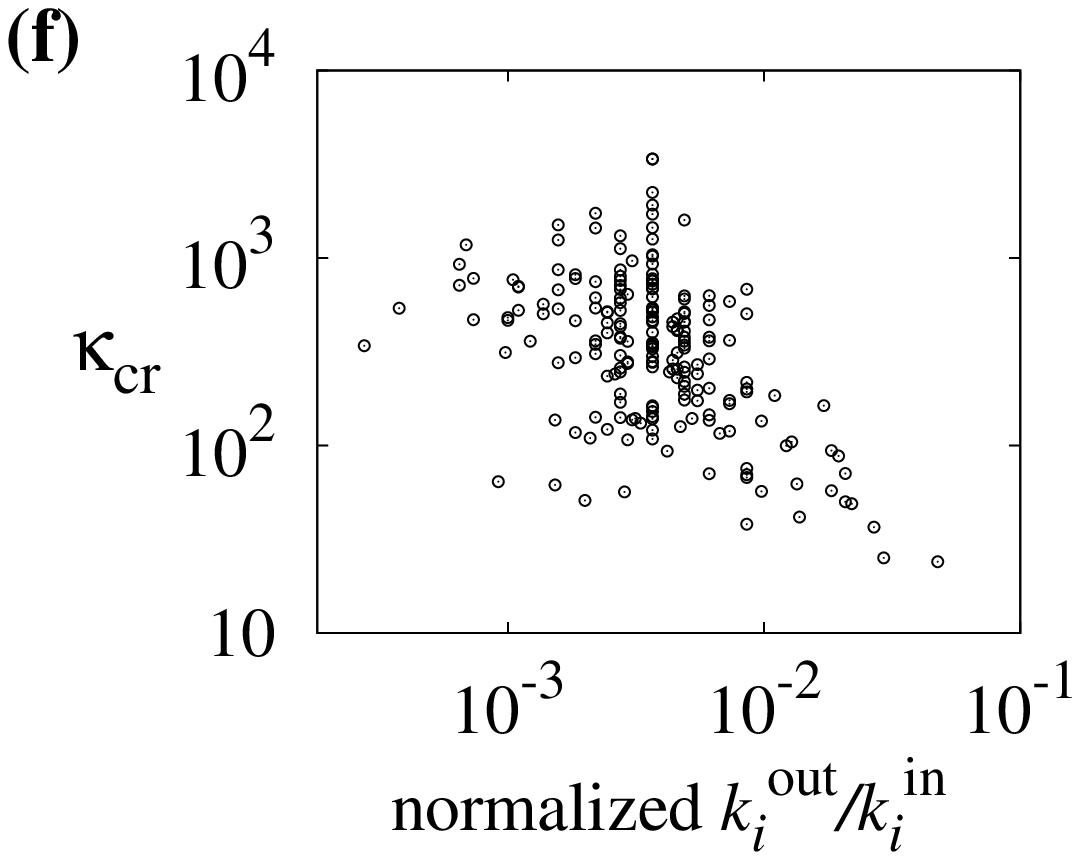}
\includegraphics[width=6cm]{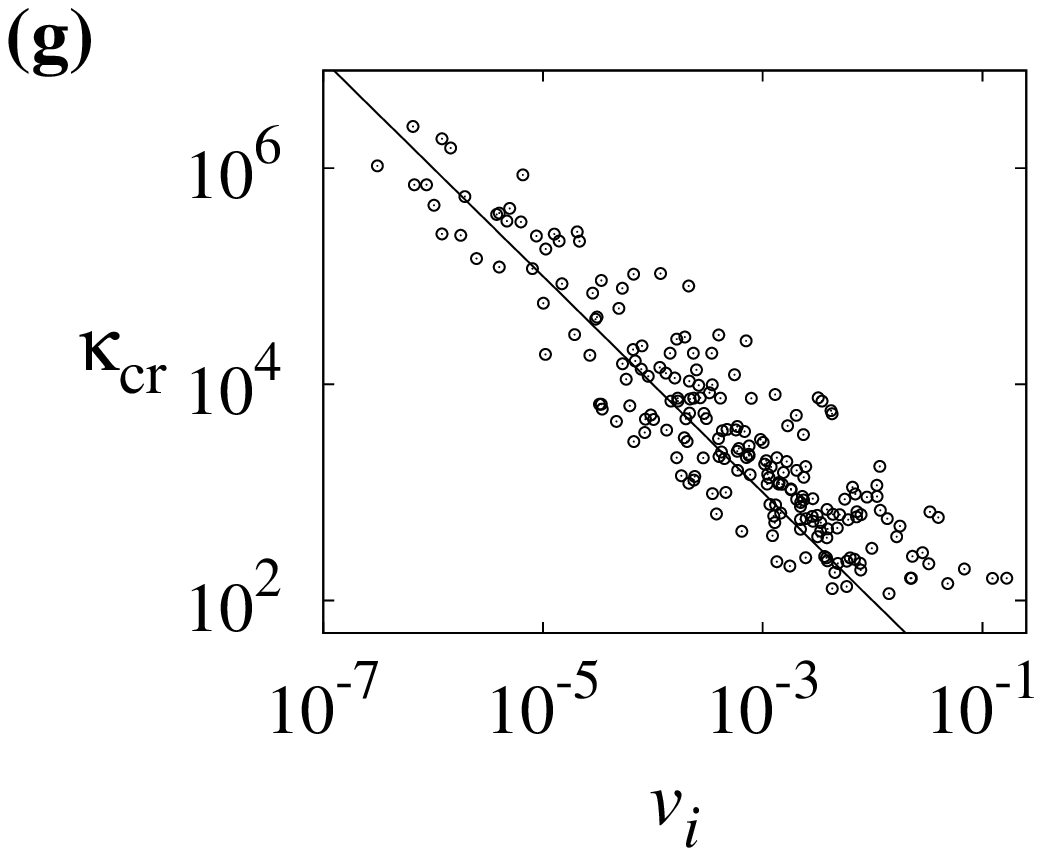}
\includegraphics[width=6cm]{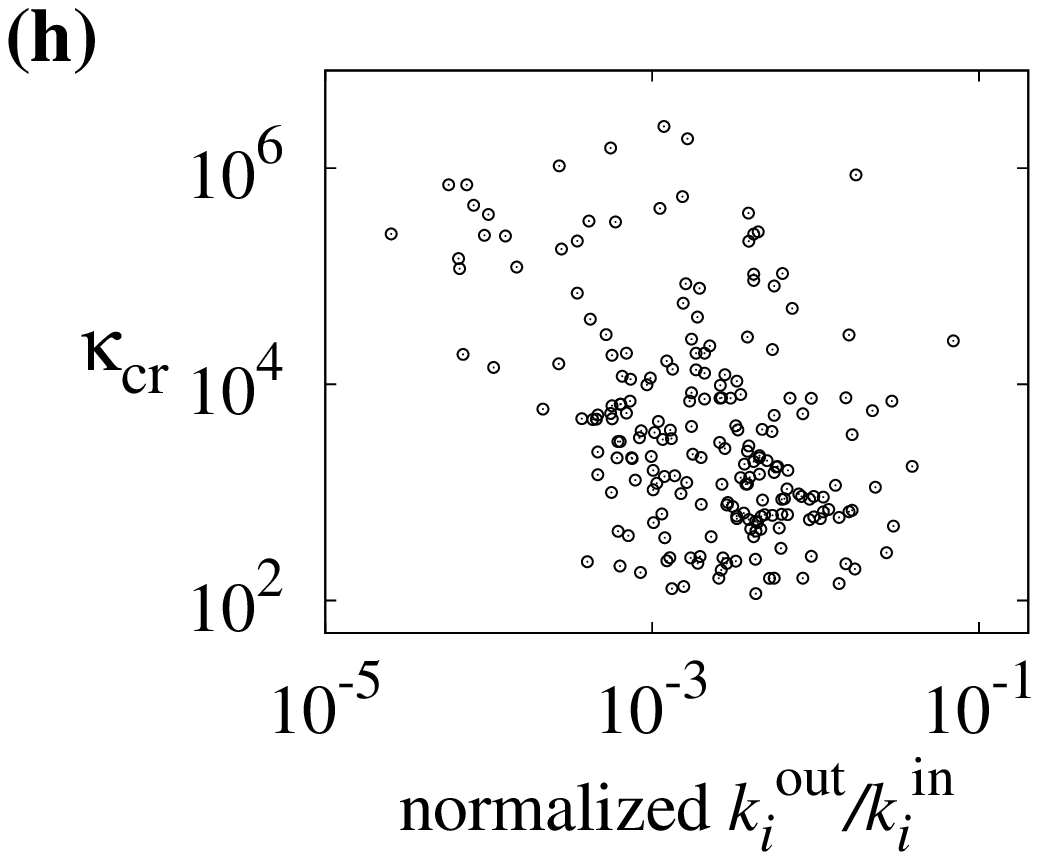}
\caption{Relationships between $\kappa_{\rm cr}$ and $v_i$. (a, b)
Directionally biased random network with $N=200$ and weight of
parallel links equal to $\epsilon$. (c, d) Directionally biased random
network with $N=200$ and weight of parallel links equal to unity. (e,
f) Directed scale-free network with $N=200$.  (g, h)
\textit{C.~elegans} neural network with $N=237$.  The data are plotted
against $v_i$ in (a, c, e, g) and against $(k_i^{\rm out}/k_i^{\rm
in})/\sum_{j=1}^N (k_j^{\rm out}/k_j^{\rm in})$ in (b, d, f, h). The
lines in (a, c, e, g) represent $\kappa_{\rm cr}\approx v_i^{-1}$.}
\label{fig:kcr}
\end{center}
\end{figure}

The values of $\kappa_{\rm cr}$ when the pacemaker is located at
different nodes are plotted against $v_i$ in
\FIG\ref{fig:kcr}(a). The line in the figure
represents $\kappa_{\rm cr}=
v_i^{-1}$, i.e., \EQ\eqref{eq:kcr vs
v_i}.  The numerically obtained $\kappa_{\rm cr}$ roughly matches the
theoretical one although the condition $\epsilon z\gg 1$ is violated.
The same values of $\kappa_{\rm cr}$ are plotted against the
MA estimate $v_i\approx (k_i^{\rm out}/k_i^{\rm in})/\sum_{j=1}^N
(k_j^{\rm out}/k_j^{\rm in})$ 
in \FIG\ref{fig:kcr}(b). We find a larger spread of data
in this plot as 
compared to that in 
\FIG\ref{fig:kcr}(a); $v_i$ predicts $\kappa_{\rm
cr}$ more accurately than the MA.

Next, we set the weight of the parallel link to unity, as done in
\cite{Kori06pre}.  The values of $\kappa_{\rm cr}$ for this version of
the directionally biased random
network are shown in \FIGS\ref{fig:kcr}(c) and
\ref{fig:kcr}(d).  The results are qualitatively the same as those
shown in \FIGS\ref{fig:kcr}(a) and \ref{fig:kcr}(b).
The dependence
of $\kappa_{\rm cr}$ on $v_i$ is weak in the new network
[\FIG\ref{fig:kcr}(c)] as compared to the previous network
[\FIG\ref{fig:kcr}(a)] mainly for the following reason.  Because of
the difference in the weights of parallel links, $\left<k\right>$ in
the new network is larger than that in the previous network.
Then, $\kappa_{\rm cr}^{(L)}$ is smaller for the new network,
since it is inversely proportional to the
effective link weight. We expect that 
$\kappa_{\rm cr}$ for nodes in intermediate layers
can also be explained using the same approach.

The result $\kappa_{\rm cr}\approx v_i^{-1}$ is derived for the
directionally biased random network.  Although it is not guaranteed
that this relationship holds true in other types of networks, we test
the applicability of the relation $\kappa_{\rm cr}\approx v_i^{-1}$ in
a scale-free network and a neural network.

We generate a directed scale-free network using the configuration
model \cite{reviews}. The degree distributions are independently given
for $k_i^{\rm in}$ and $k_i^{\rm out}$ by $p(k^{\rm in})\propto
k^{-\gamma_{\rm in}}$ and $p(k^{\rm out})\propto k^{-\gamma_{\rm
out}}$, respectively. We set $\gamma_{\rm in}=\gamma_{\rm out}=2.5$
and $N=200$. The minimum degree is set to 3.  The duration of a run
and the length of the traisient are equal to those in the case of the
directionally biased random networks. The values of $\kappa_{\rm cr}$
are plotted against $v_i$ and the MA in \FIGS\ref{fig:kcr}(e) and
\ref{fig:kcr}(f), respectively.  The relation $\kappa_{\rm cr}\approx
v_i^{-1}$ fits the data reasonably well, even though the scale-free
network is not a directionally biased random network.  In contrast,
$k_i^{\rm out}/k_i^{\rm in}$ poorly predicts $\kappa_{\rm cr}$, as
shown in \FIG\ref{fig:kcr}(f).

We next examine the \textit{C.~elegans} neural network
\cite{Chen06pnas,wormatlas} based on chemical synapses, which serve as
directed links. The network has the largest strongly connected
component with $N=237$ nodes. The number of synapses from neuron $i$
to neuron $j$ defines $w_{ij}$.  We set $T=2.5\times 10^7/\kappa$ to
appropriately exclude the transient. The values of $\kappa_{\rm cr}$
are plotted against $v_i$ and the MA in \FIGS\ref{fig:kcr}(g) and
\ref{fig:kcr}(h), respectively.  The values of $\kappa_{\rm cr}$
exceeding $10^7$ are not plotted because direct numerical simulations
need too much time. Similar to the case of scale-free networks, $v_i$
predicts $\kappa_{\rm cr}$ better than $k_i^{\rm out}/k_i^{\rm in}$
does.
 
\section{Conclusions}

In this paper, we have analyzed the centrality measure called the
influence in various networks. The influence extracts the magnitude
with which a node controls or impacts the entire network along
directed links.  We have analytically shown that the source of the
shortcut in a directed version of the Watts-Strogatz small-world
network and the root node in hierarchical networks have large
influence values.  This is not accurately predicted if we approximate
the influence of a node by its degree. Although by definition, the
influence is based on local connectivity, the global structure of
networks does affect the influence values. We also analyzed the effect
of the location of a pacemaker on the capability of entrainment in a
system of coupled phase oscillators.  The pacemaker located at a node
with a large influence value entrains the other oscillators relatively
easily.

In the analysis of some model networks, including the Watts-Strogatz
small-world network, we used the method based on the enumeration of
directed spanning trees. This method can be applied to the estimation
of the PageRank of nodes, because the PageRank can be mapped to the
influence if we reverse links and rescale the link weight
\cite{MKK09-1,Masuda09njp}.  Application of our results to other
centrality measures for directed networks is warranted for future
study.

\begin{acknowledgments}
N.M. acknowledges the support provided by
the Grants-in-Aid for Scientific Research
(Grants No. 20760258 and No. 20540382) from MEXT, Japan.
\end{acknowledgments}

\newpage
\clearpage

\clearpage

\clearpage

\clearpage

\clearpage

\clearpage

\clearpage

\clearpage

\clearpage

\end{document}